\documentclass[12pt]{iopart}

\usepackage{graphicx}

\usepackage{iopams}
\usepackage{amsfonts}

\begin{document}

\title[Two interacting particles]
{Poincar\'e invariant interaction between two Dirac particles }

\author{Juan Barandiaran}
\address{Telecommunication Engineer, Bilbao, Spain}
\ead{barandiaran.juan@gmail.com}
\author{Mart\'{\i}n Rivas}
\address{Theoretical Physics Department, The University of the Basque Country,\\ 
Bilbao, Spain}
\ead{martin.rivas@ehu.eus}

\begin{abstract}
The spinning electron-electron interaction is described in classical terms by means of two possible classical interactions: The instantaneous Coulomb interaction between the charge centers of both particles and the Poincar\'e invariant interaction developed in a previous work. The numerical integrations are performed with several Mathematica notebooks that are available for the interested readers in the reference Section. One difference of these interactions is that the Poincar\'e invariant interaction does not satisfy the action-reaction principle in the synchronous description and, therefore, there is no conservation of the mechanical linear momentum. It is the total linear momentum of the system what is conserved. 

In this synchronous description the interaction is not mediated by the retarded fields but is described in terms of the instantaneous positions, velocities and accelerations of the center of charge of both particles. In the Poincar\'e invariant description the net binding force that holds linked two Dirac particles is stronger than in the Coulomb case, thus forming a stable spin 1 system of 2 Dirac particles. This bosonic state of spin 1 does not correspond to a Cooper pair because the separation between the centers of mass of the Dirac particles is below Compton's wavelength, smaller than the correlation distance of the Cooper pair. Since the Poincar\'e invariant interaction is relativistically invariant it can be used for analyzing high energy scattering processes.
\end{abstract}

\hspace{1.2cm}{\small\bf Keywords:} {Spinning electron; Dirac particle; Coulomb interaction; Poincar\'e invariant interaction; spinning electrons bound state.}

\section{Introduction}
\label{intro}
We are going to analyze the electron-electron interaction by making use of a classical model of spinning electron, that is described in section {\bf\ref{clasicalelectron}}. This model satisfies Dirac's equation when quantized and we call it the {\it classical Dirac particle}. We are going to describe the interaction of two Dirac particles by means of the instantaneous Coulomb interaction between both particles and to compare these results with the Poincar\'e invariant interaction obtained in a previous work \cite{InvariantLag}. 

These two Lagrangians are described in section {\bf\ref{interaction}}. The main feature of the classical spinning electron is that we only need to describe the evolution of a single point, the center of charge, where the external force is defined, and which satisfies a system of fourth-order differential equations. All observables are expressed in terms of this point and their time derivatives.

The spinning electron has also another characteristic point, the center of mass, which is a different point than the center of charge and satisfies Newton-type dynamical equations in terms of the external force evaluated at the position of the center of charge. In section {\bf\ref{conservation}} we analyze the conservation laws of both Lagrangians. The center of charge is moving at the speed of light and the complete dynamical equations for the evolution of the center of mass and center of charge of both particles can be expressed in terms of dimensionless classical variables, where the only dimensionless and invariant parameter which controls the intensity of the interaction, is the fine structure constant.  The whole formalism can be described in terms of a natural system of units (Section {\bf\ref{Natural}}). In sections {\bf\ref{Coulomb}} and {\bf\ref{Poincare}} we compute the boundary initial conditions 
for the two particles, in natural units, under the Coulomb and Poincar\'e interactions, respectively. The difference lies on the non-conservation of the mechanical linear momentum of the particles in the Poincar\'e case, because it is the total linear momentum what is conserved, as analyzed in section {\bf\ref{conservation}}. In section {\bf\ref{PoincarInter}} we analyze examples under the Poincar\'e invariant interaction. In section {\bf\ref{CoulombPoincarInter}} we describe different examples under the Coulomb and Poincar\'e invariant interaction, respectively, to analyze the difference between both interactions. In section {\bf\ref{Pairing}} we justify in classical terms why two Dirac particles can form a bound stable state of spin 1. We end the work with some general analysis of the interactions of two Dirac particles in the section {\bf\ref{Conclusions}}.

\section{Classical Dirac particle}
\label{clasicalelectron}

The model of a classical elementary spinning particle that we are going to use in this work, has been obtained through a general formalism \cite{Rivasbook}, that is based on the following three fundamental principles: Restricted Relativity Principle, Variational Principle and Atomic Principle. 

The definition of a classical elementary particle lies on the Atomic principle \cite{atomic}. The idea is that an elementary particle does not have internal excited states in the sense that any interaction, if it does not annihilate the particle, does not modify its internal structure. If an inertial observer describes the initial state of the elementary particle in the Lagrangian description through a set of variables $(x_1,\ldots,x_n)$ and the dynamics changes this state to  another $(y_1,\ldots,y_n)$, if the internal structure has not been modified, then it is possible to find another inertial observer who describes this new state of the particle with exactly the same values of the variables as of the previous state. This means that there must exist a Poincar\'e transformation $g$, between both inertial observers, in such a way that the above values are transformed into each other: 
\[
(y_1,\ldots,y_n)=g\,(x_1,\ldots,x_n).
\]
This relation must be valid for any pair of states of any elementary particle.
The Atomic Principle restricts the classical variables of the variational description of an elementary particle, to span a homogeneous space of the Poincar\'e group \cite{Rivasbook}. Then the initial and final states of the Lagrangian description are described by as many variables as those of any Poincar\'e group element, and with the same geometrical or physical meaning like the group parameters. Because the Poincar\'e group is a ten-parameter Lie group the initial state is characterized at most by ten variables $x_i\equiv(t,{\bi r},{\bi u},\balpha)$, that are interpreted as the time $t$, the position of a single point ${\bi r}$, the velocity of this point ${\bi u}$,
and finally the orientation $\balpha\in SO(3)$, of a comoving Cartesian frame attached to the point ${\bi r}$. The same variables, with different values, for the final point $x_f$ of the Lagrangian evolution. 
If we restrict ourselves to analyze a mechanical system where the initial state is given by the variables $x_i\equiv(t,{\bi r})$, which span a homogeneous space of the Poincar\'e group, we are describing also an elementary particle: the spinless point particle. It is in terms of the extra variables $({\bi u},\balpha)$, that we will be able to describe the spin structure.

Among the models of spinning particles this formalism predicts, the only one that satisfies Dirac's equation when quantized \cite{RivasDirac}, is that model in which the point ${\bi r}$, is moving at the speed of light $u=c$. Since the Lagrangian also depends  on the next order time derivative of the above boundary variables, the Lagrangian must depend on the acceleration of the point ${\bi a}$, and on the angular velocity $\bomega$. The point ${\bi r}$, satisfies therefore  a system of fourth-order differential equations and, being the only point where the external fields are defined, we interpret the point ${\bi r}$, as the location of the center of charge (CC) of the particle.

We must remember that according to Frenet-Serret formalism, the family of continuous and differentiable trajectories of points with continuous and differentiable curvature and torsion
in three-dimensional space, satisfy a system of fourth-order differential equations.

\begin{figure}[!hbtp]\centering%
\includegraphics[width=7cm]{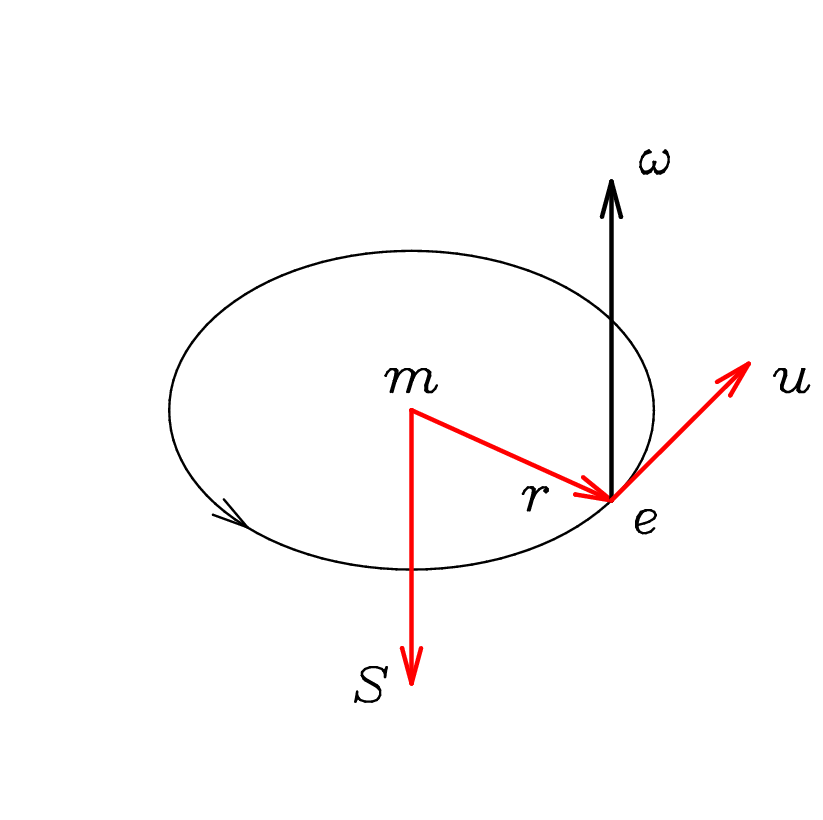}
\caption{This model represents the circular motion, at the speed of light, of the center
of charge of the electron in the center of mass frame, as described by the dynamical equation (\ref{constS}). The center of mass is always a different point than the center of charge. The spin ${\bi S}$ has the opposite direction to the angular velocity $\bomega$.
The radius of this motion is $R_0=\hbar/2mc$, in this frame.
The angular velocity is $\omega=2mc^2/\hbar$.} 
\label{fig1:elecCM}
\end{figure}

This model is depicted in figure {\bf\ref{fig1:elecCM}} for the center of mass observer. It has a center of charge (CC), ${\bi r}$, that moves at the speed of light $c$, around the center of mass (CM), ${\bi q}$, that is at rest in this frame, and that is a different point than the CC. In the free motion the CC describes a circle of radius $R_0$, contained in a plane orthogonal to the spin. We will call to this model from now on, a {\bf classical Dirac particle}.  

The position ${\bi r}$ of the CC, satisfies a system of fourth-order differential equations
\cite{RivasDynamics}, that can be separated into a system of second-order differential equations for the CC and CM positions, where the CM position ${\bi q}$, is defined in terms of the motion of the CC, ${\bi r}$, by 
\begin{equation}
{\bi q}={\bi r}+\left(\frac{c^2-{\bi v}\cdot{\bi u}}{(d^2{\bi r}/dt^2)^2}\right)\frac{d^2{\bi r}}{dt^2},
\label{defposCM}
\end{equation}
where
${\bi v}={d{\bi q}}/{dt}$, and ${\bi u}={d{\bi r}}/{dt}$ and $v<c$ and $u=c$.
The energy $H$, and the linear momentum ${\bi p}$, are finally written, as in the case of the point particle model, in terms of the center of mass velocity ${\bi v}$, as:
\begin{equation}
H=\gamma(v)mc^2,\quad {\bi p}=\frac{H}{c^2}{\bi v}=\gamma(v)m{\bi v},\quad \gamma(v)=(1-v^2/c^2)^{-1/2}.
\label{defHyp}
\end{equation}

The relativistic differential equations in the presence of an external electromagnetic field ${\bi E}(t,{\bi r})$ and ${\bi B}(t,{\bi r})$, defined at the CC position ${\bi r}$, and in any arbitrary inertial reference frame are \cite{RivasDynamics}: 
\begin{eqnarray}
\frac{d^2{\bi q}}{dt^2}&=&\frac{e}{m\gamma(v)}\left[{\bi E}+{\bi u}\times{\bi B}-\frac{1}{c^2}{\bi v}\left(\left[{\bi E}
+{\bi u}\times{\bi B}\right]\cdot{\bi v}\right)\right],\label{eq:d2qdt2}\\
\frac{d^2{\bi r}}{dt^2}&=&\frac{c^2-{\bi v}\cdot{\bi u}}{({\bi q}-{\bi r})^2}\left({\bi q}-{\bi r}\right),\label{eq:d2rdt2}
 \end{eqnarray}
with the constraint $|{\bi u}|=c$ and $|{\bi v}|<c$. 
The Lagrangian which gives rise to these dynamical equations is
\[
L=L_0({\bi u},{\bi a},\bomega)+L_{em}(t,{\bi r},{\bi u}),\quad L_{em}=-eA_0(t,{\bi r})+e{\bi u}\cdot{\bi A}(t,{\bi r})
\]
where $L_0$ is the free Lagrangian, $e$ is the electric charge of the particle and $A_0$ and ${\bi A}$, are, respectively the scalar and vector potentials defined at the CC of the particle and ${\bi u}$ is the CC velocity.

The free Lagrangian $L_0({\bi u},{\bi a},\bomega)$, is translation invariant and therefore it is a function of the velocity, acceleration and angular velocity of the CC. The fourth-order Euler-Lagrange equations of the position variables are\footnote{\footnotesize{We represent 3D vector observables in boldface. Expressions like ${\bi a}=\partial L/\partial {\bi b}$, have to be interpreted throughout this work as $a_i=\partial L/\partial b_i$, $i=1,2,3$.}}
\[
\frac{\partial L_0}{\partial{\bi r}}-\frac{d}{dt}\left(\frac{\partial L_0}{\partial{\bi u}}\right)+
\frac{d^2}{dt^2}\left(\frac{\partial L_0}{\partial{\bi a}}\right)+\frac{\partial L_{em}}{\partial{\bi r}}
-\frac{d}{dt}\left(\frac{\partial L_{em}}{\partial{\bi u}}\right)=0.
\]
Since $L_0$ is independent of ${\bi r}$, the above dynamical equation reads:
\[
-\frac{d}{dt}\left[\frac{\partial L_0}{\partial{\bi u}}-
\frac{d}{dt}\left(\frac{\partial L_0}{\partial{\bi a}}\right)\right]-e(\nabla A_0+\partial{\bi A}/\partial t)+e{\bi u}\times(\nabla\times{\bi A})=0,
\]
and the term between the squared brackets represents the mechanical linear momentum of the particle ${\bi p}$.
This equation is
\[
\frac{d{\bi p}}{dt}=e\left({\bi E}+{\bi u}\times{\bi B}\right),
\]
and corresponds to equation (\ref{eq:d2qdt2}) 
where ${\bi u}$ is the velocity of the CC, which appears in the magnetic force. From ${\bi p}$  defined in (\ref{defHyp}) its time derivative is expressed in terms of $d^2{\bi q}/dt^2$, and writing on the left hand side this derivative we get (\ref{eq:d2qdt2}).
Equation (\ref{eq:d2rdt2}) is equation (\ref{defposCM}) when writing on the left hand side the acceleration of the CC. Taking the second order derivative of ${\bi q}$ in (\ref{defposCM}) and replacing it by (\ref{eq:d2qdt2}) we recover \cite{RivasDynamics} the fourth-order system of differential equations for the CC position ${\bi r}$, which is where the external fields ${\bi E}(t,{\bi r})$ and ${\bi B}(t,{\bi r})$, are defined. 

The point particle free Lagrangian is a function $L_0({\bi u})$ of the velocity variable of the
point ${\bi r}$ and is independent of the time $t$ and position ${\bi r}$. The angular momentum with respect to the origin is obtained by applying the rules of Noether's theorem  and gives ${\bi J}={\bi r}\times{\bi p}$, where the linear momentum is obtained from the Lagrangian as ${\bi p}=\partial L_0/\partial{\bi u}$. 
For the spinning particle, the free Lagrangian is a function $L_0({\bi u},{\bi a},\bomega)$. Nother's theorem defines the angular momentum with respect to the origin as ${\bi J}={\bi r}\times{\bi p}+{\bi u}\times{\bi U}+{\bi W}$, where ${\bi U}=\partial L_0/\partial{\bi a}$, ${\bi W}=\partial L_0/\partial{\bomega}$ and ${\bi p}=\partial L_0/\partial{\bi u}-d{\bi U}/dt$. Therefore ${\bi J}={\bi r}\times{\bi p}+{\bi S}$, where the function ${\bi S}={\bi u}\times{\bi U}+{\bi W}$, represents the angular momentum with respect to the point ${\bi r}$. It is the spin with respect to the CC. The additional variables ${\bi a}$ and $\bomega$ in the Lagrangian, that define the observables ${\bi U}$ and ${\bi W}$, provide the structure of the spin when we compare it with the point particle \cite{Rivasbook}.

Since this model has two characteristic points, the formalism defines also the angular momentum (spin) of the particle with respect to the point ${\bi q}$. It is found, by the above Noether analysis of the free particle, that the expression of the spin ${\bi S}$ with respect to the CC, ${\bi r}$, can be written as:
\begin{equation}
{\bi S}=-\gamma(v)m({\bi r}-{\bi q})\times{\bi u},
\label{eq:spinCC}
\end{equation}
while the spin with respect to the CM, ${\bi q}$, is defined by:
\begin{equation}
{\bi S}_{CM}={\bi S}+{(\bi r}-{\bi q})\times{\bi p}=-\gamma(v)m({\bi r}-{\bi q})\times({\bi u}-{\bi v}).
\label{eq:spinCM}
\end{equation}
The spins (\ref{eq:spinCC}) and (\ref{eq:spinCM}) are finally expressed in terms of the instantaneous separation between both centers and of the velocities of both points, while it is only the velocity of the CM ${\bi v}$, which appears in the definition of $H$ and ${\bi p}$, (\ref{defHyp}).

The total angular momentum ${\bi J}$ of the particle with respect to the origin of any arbitrary inertial frame, can be written in two different ways as:
\[
{\bi J}={\bi S}+{\bi r}\times{\bi p}={\bi S}_{CM}+{\bi q}\times{\bi p}.
\]
Both spins satisfy different dynamical equations. If the particle is free, ${\bi v}$ is constant and $d{\bi J}/dt=0$, and $d{\bi p}/dt=0$, leads to
\[
\frac{d{\bi S}}{dt}={\bi p}\times{\bi u},\quad \frac{d{\bi S}_{CM}}{dt}=0.
\]
The CM spin ${\bi S}_{CM}$, is conserved while the spin with respect to the CC, ${\bi S}$, satisfies the same dynamical equation as Dirac's spin operator in the quantum case. Its time derivative is always orthogonal to the direction of the linear momentum. We must remark that in the quantum Dirac's analysis, the only variable that defines the position of the electron ${\bi r}$, is contained in Dirac's spinor $\psi(t,{\bi r})$,  it is moving at the speed $c$ and it is also the point where the external electromagnetic field $A_\mu(t,{\bi r})$ is defined. Clearly, the spin with respect to the CC, ${\bi S}$, represents the classical equivalent of Dirac's spin operator.

If the particle is free, equation (\ref{eq:d2qdt2}) implies that the CM ${\bi q}$, moves along a straight line at a constant velocity ${\bi v}$. For the center of mass observer ${\bi q}={\bi v}=0$, and both spins take the same value in this frame:
\begin{equation}
{\bi S}={\bi S}_{CM}=-m{\bi r}\times{\bi u},
\label{constS}
\end{equation}
and this constant spin is orthogonal to the position and velocity of the CC. Since $u=c$ is constant, the acceleration is also orthogonal to the velocity and satisfies in this frame the differential equation (\ref{eq:d2rdt2}):
\[
\frac{d^2{\bi r}}{dt^2}=-\frac{c^2}{r^2}\,{\bi r}.
\]
Therefore, ${\bi r}$ is also orthogonal to the velocity and if we take the constant spin ${\bi S}$ along the negative direction of $OZ$ axis, the equation (\ref{constS}) implies that the motion of the CC in the CM frame is the one depicted in the figure {\bf\ref{fig1:elecCM}}. We see that the spin ${\bi S}$ has the opposite direction to the angular velocity of a local cartesian frame attached to the point ${\bi r}$. It seems that the spin has an anti-orbital orientation with respect to the trajectory of the point ${\bi r}$, but point ${\bi r}$ does not represent the location of any point mass.

The energy can also be written in the form:
\begin{equation}
H={\bi p}\cdot{\bi u}+\frac{1}{c^2}{\bi S}\cdot\left(\frac{d{\bi u}}{dt}\times{\bi u}\right),
\label{eq:DiracH}
\end{equation}
which contains two terms: the first term, proportional to the linear momentum, represents the translational energy, while the second, proportional to the spin and to the motion of the CC (zitterbewegung), represents the rotational energy which never vanishes. If we introduce in (\ref{eq:DiracH}) the expressions of ${\bi p}$ and ${\bi S}$, given in (\ref{defHyp}) and (\ref{eq:spinCC}), respectively, we get $H=\gamma(v)mc^2$. It is this linear expression in terms of $H$ and ${\bi p}$, which will give rise to Dirac's Hamiltonian when quantizing the model \cite{RivasDirac}. In the quantum description $H=i\hbar\partial/\partial t$ and ${\bi p}=-i\hbar\nabla$ and (\ref{eq:DiracH}) is transformed into Dirac equation.
 The energy contains two terms: the first term, proportional to the linear momentum, represents the translational energy, while the second, proportional to the spin and angular velocity and to the motion of the CC (zitterbewegung), and independent of the CM velocity, represents the rotational energy $H_{rot}$ which is positive definite, because the spin ${\bi S}$ has the opposite direction to the angular velocity $\bomega_p$, and never vanishes.

Although the formalism is based on the existence of a Lagrangian, all the above expressions of the different observables $H$, ${\bi p}$, ${\bi q}$ and ${\bi S}$, for the free Dirac particle in terms of the position of the the center of charge and their derivatives, have been obtained without the explicit knowledge of the free Lagrangian $L_0$. They are obtained by the Noether analysis applied to the assumed Poincar\'e invariance of the dynamical equations \cite{Rivasbook}, as required by the Restricted Relativity Principle. The conserved Noether observables are not a function of the Lagrangian but they depend explicitely on the partial derivatives of the Lagrangian with respect to the derivatives of the kinematical variables, and these partial derivatives are the observables which have a clear relationship and interpretation with the above conserved quantities.

The dynamical equations (\ref{eq:d2qdt2}) and (\ref{eq:d2rdt2}) have been obtained \cite{Rivasbook} by invariance arguments applied to a Lagrangian description of a mechanical system whose boundary variables manifold of the Lagrangian formalism spans a homogeneous space of the Poincar\'e group, with the constraint $u=c$. For people not interested in the way how these equations were obtained, they can consider the above equations (\ref{eq:d2qdt2}) and (\ref{eq:d2rdt2}) as the fundamental equations that describe the evolution of the CC and CM of the Dirac particle in an external electromagnetic field. 

Although this formalism is not written in covariant form in terms of four-vectors, it is nevertheless an invariant formalism in the sense that the expressions of the dynamical equations and of the different observables are written in the same form in each inertial reference frame, where the time variable is the time coordinate in that frame.

\subsection{The angular velocity}
\label{angular}

The angular velocity corresponds to the rotation of the comoving Cartesian frame linked to the point ${\bi r}$. But this frame is completely arbitrary. Once the CC is moving, its dynamics defines the acceleration ${\bi a}$, orthogonal to the velocity ${\bi u}$, and therefore
these two orthogonal vectors with the vector ${\bi u}\times{\bi a}$ define an orthogonal comoving frame (the Frenet-Serret frame). Since ${\bi u}$ is a vector of constant absolute value, its time derivative the acceleration ${\bi a}$, is orthogonal to it and it can be written as
\[
{\bi a}={\bomega}\times{\bi u}.
\]
The cross product of this expression with ${\bi u}$ gives:
\[
{\bi a}\times{\bi u}={\bi u}({\bomega}\cdot{\bi u})-c^2\bomega.
\]
The first term is the component of the angular velocity along the velocity ${\bi u}$, $\bomega_u$ times $c^2$, so that $\bomega-\bomega_u=\bomega_p=({\bi u}\times{\bi a})/c^2$. Then the perpendicular component of the angular velocity $\bomega_p$ to the plane defined by the vectors ${\bi u}$ and ${\bi a}$, can be expressed as
\begin{equation}
\bomega_p=\frac{1}{c^2}({\bi u}\times{\bi a})=\frac{1}{c^2}\left(\frac{d{\bi r}}{dt}\times{\frac{d^2{\bi r}}{dt^2}}\right).
\label{omegaperp}
\end{equation}
This perpendicular component has the opposite direction than the CC spin since from (\ref{eq:spinCC}) and (\ref{eq:d2rdt2}) $({\bi r}-{\bi q})\simeq -{\bi a}$. In general the CM spin and the CC spin are not along the angular velocity $\bomega$. This perpendicular component is responsible for the curvature of the trajectory, while the $\bomega_u$ is related to the torsion of the trajectory. In the center of mass frame the trajectory of the point ${\bi r}$ is flat, there is no torsion and thus $\bomega_u=0$.

The component $\bomega_u$ produces the torsion of the trajectory and depends on the third orden derivative of the vector ${\bi r}$. This component of the angular velocity is
\begin{equation}
\bomega_u=\frac{1}{c^2a^2}\left(({\bi u}\times{\bi a})\cdot\frac{d{\bi a}}{dt}\right){\bi u}=\frac{1}{c^2a^2}\left[\left(\frac{d{\bi r}}{dt}\times{\frac{d^2{\bi r}}{dt^2}}\right)\cdot\frac{d^3{\bi r}}{dt^3}\right]\frac{d{\bi r}}{dt}.
\label{omegau}
\end{equation}

The evolution of the angular velocity is completely determined by the motion of the CC. The evolution of the CC determines both the trajectories of the CC and CM and the rotation of the body frame (Frenet-Serret frame) attached to the point ${\bi r}$.

We can find an alternative expression for the angular velocity $\bomega_u$ if we use the dynamical equation (\ref{eq:d2rdt2}). Taking the next order time derivative we have:
\[
\frac{d^3{\bi r}}{dt^3}=\frac{(-{\bi a}_{CM}\cdot{\bi u}-{\bi v}\cdot{\bi a})({\bi q}-{\bi r})^2-2({\bi q}-{\bi r})\cdot({\bi v}-{\bi u})(c^2-{\bi v}\cdot{\bi u})}{({\bi q}-{\bi r})^4}\,({\bi q}-{\bi r})+
\]
\[
+\frac{c^2-{\bi v}\cdot{\bi u}}{({\bi q}-{\bi r})^2}\,({\bi v}-{\bi u}).
\]
The term in squared brackets in (\ref{omegau})
\[
\left(\frac{d{\bi r}}{dt}\times{\frac{d^2{\bi r}}{dt^2}}\right)\cdot\frac{d^3{\bi r}}{dt^3}=\left(\frac{d^2{\bi r}}{dt^2}\times{\frac{d^3{\bi r}}{dt^3}}\right)\cdot\frac{d{\bi r}}{dt}
=\left({\bi a}\times{\frac{d^3{\bi r}}{dt^3}}\right)\cdot{\bi u},
\]
and since from (\ref{eq:d2rdt2}) ${\bi a}\simeq ({\bi q}-{\bi r})$, the last term in this expression is 
\[
\frac{c^2-{\bi v}\cdot{\bi u}}{({\bi q}-{\bi r})^2}\,({\bi a}\times({\bi v}-{\bi u}))\cdot{\bi u}.
\]
Similarly the term
\[
({\bi a}\times({\bi v}-{\bi u}))\cdot{\bi u}=(({\bi v}-{\bi u})\times{\bi u})\cdot{\bi a}=({\bi v}\times{\bi u})\cdot{\bi a}=({\bi u}\times{\bi a})\cdot{\bi v}.
\]
Collecting terms, instead of the expression (\ref{omegau}) we obtain
\begin{equation}
\bomega_u=\frac{1}{c^2a^2}\frac{c^2-{\bi v}\cdot{\bi u}}{({\bi q}-{\bi r})^2}\left[({\bi u}\times{\bi a})\cdot{\bi v}\right]{\bi u}=\left[\frac{\bomega_p\cdot{\bi v}}{c^2-{\bi v}\cdot{\bi u}}\right]{\bi u}.
\label{omegaunew}
\end{equation}
If the CM velocity ${\bi v}=0$, or along to the zitterbewegung plane, $\bomega_p\cdot{\bi v}=0$ and thus $\bomega_u=0$, the CC trajectory is a flat curve and there is no torsion. If $\bomega_p$ and ${\bi v}$ form an angle $\alpha<\pi/2$, $\bomega_u$ has the direction of ${\bi u}$ and the opposite direction if $\alpha>\pi/2$. 

The evolution of the angular velocity is determined by the motion of the CC.
In this way, the above rotational part of the Hamiltonian (\ref{eq:DiracH}) is just $-{\bi S}\cdot{\bomega}>0$,
because the spin has the direction opposite to the angular velocity of this frame, and never vanishes. The classical equivalent to Dirac's Hamiltonian (\ref{eq:DiracH}) can also be written as $H={\bi p}\cdot{\bi u}-{\bi S}\cdot{\bomega}$, where the translation and rotation energies are more evident.

In the Lagrangian description we started with a mechanical system of 6 degrees of freedom. Three represent the position of a point ${\bi r}$, and another three $\balpha\in SO(3)$, that describe the orientation of a comoving frame attached to the point ${\bi r}$. The dynamics has established a constraint such that the degrees of freedom $\balpha$ can be related to the Frenet-Serret frame of the motion of ${\bi r}$, that satisfies a system of fourth-order differential equations. This point and their derivatives, completely determine the definition of any other observable. The electron description has been reduced to the evolution of a single point, the center of charge.

\section{Interaction of two Dirac particles}
\label{interaction}

\subsection{Coulomb interaction}
We are going to calculate the interaction of two equal Dirac particles of charge $e$ and mass $m$ by means of the instantaneous Coulomb interaction. The Lagrangian which describes this system is
\begin{equation}
L=L_1+L_2+L_C,\quad L_C=-\frac{e^2}{4\pi\epsilon_0}\frac{1}{|{\bi r}_1-{\bi r}_2|},
\label{eq:LagrangianCoul}
\end{equation}
where $L_1$ and $L_2$ are the free Lagrangians of the two particles. The Lagrangian $L_C$ represents the instantaneous Coulomb potential energy between the two charges. It depends on the relative distance between the CC's of both particles. The free Lagrangians $L_a({\bi u}_a,{\bi a}_a)$, are independent of the time and position of the corresponding particle and depend only the velocity and acceleration of the CC of each particle. The CC of each free particle ${\bi r}_a$, satisfies a system of fourth-order ordinary differential equations as described in section {\bf\ref{clasicalelectron}}, which can be separated into a system of second-order differential equations for the CC's and CM's of each particle. This total Lagrangian has dimensions of energy and we are in the temporal description of the evolution.

The dynamical equations to be solved in the synchronous description at any
arbitrary inertial reference frame are \cite{RivasDynamics}: 
\begin{eqnarray}
\frac{d^2{\bi q_a}}{dt^2}&=&\frac{1}{m\gamma(v_a)}\left[{\bi F}_a-\frac{1}{c^2}{\bi v}_a\left({\bi F}_a\cdot{\bi v}_a\right)\right],\label{eq:Could2qdt2}\\
\frac{d^2{\bi r}_a}{dt^2}&=&\frac{c^2-{\bi v}_a\cdot{\bi u}_a}{({\bi q}_a-{\bi r}_a)^2}\left({\bi q}_a-{\bi r}_a\right),\quad a=1,2,\label{eq:Could2rdt2}
 \end{eqnarray}
where ${\bi F}_a$, $a=1,2$ is the Coulomb force acting on the particle $a$. The force on particle 1 is
\begin{equation}
{\bi F}_1=\frac{\partial L_C}{\partial {\bi r}_1}=\frac{e^2}{4\pi\epsilon_0}\frac{{\bi r}_1-{\bi r}_2}{|{\bi r}_1-{\bi r}_2|^3}=-\frac{\partial L_C}{\partial {\bi r}_2}=-{\bi F}_2,
\label{CoulombForce}
\end{equation}
opposite to the force acting on the particle 2, ${\bi F}_2$.

\subsection{Poincar\'e invariant interaction}

We are going to calculate  as before the interaction of two equal Dirac particles of charge $e$ and mass $m$ by means of the Poincar\'e invariant interaction obtained in \cite{InvariantLag}. The Lagrangian which describes this system in the synchronous description is
\begin{equation}
L=L_1+L_2+L_P,\quad L_P=-\frac{e^2}{4\pi\epsilon_0c}\frac{\sqrt{c^2-{\bi u}_1\cdot{\bi u}_2}}{|{\bi r}_1-{\bi r}_2|},
\label{eq:LagrangianPoin}
\end{equation}
where as before $L_1$ and $L_2$ are the free Lagrangians of the two Dirac particles. This total Lagrangian has dimensions of energy and we are in the temporal description of the evolution.

\begin{quotation}{\footnotesize\noindent{\bf Poincar\'e Invariant interaction Lagrangian}\\
Let  $x_a^\mu\equiv (ct_a,{\bi r}_a)$ and $\dot{x}_a^\mu\equiv (c\dot{t}_a,\dot{\bi r}_a)$, $a=1,2$, be the four-vectors which define the space-time coordinates and four-velocities of both particles, respectively.
The overdot represents the derivative with respect to some invariant and dimensionless evolution parameter $\tau$.
The variational integral in terms of some arbitrary evolution parameter $\tau$ can be rewritten as:
\[
\int_{t_1}^{t_2}L\,dt=\int_{\tau_1}^{\tau_2}L\,\dot{t}(\tau)d\tau=\int_{\tau_1}^{\tau_2}\widetilde{L}\,d\tau,\quad \widetilde{L}=L\dot{t}(\tau),
\]
and $\widetilde{L}$, is a homogeneous function of first degree in the $\tau$-derivatives of the boundary variables \cite{Rivasbook}. If the evolution parameter $\tau$ is dimensionless, this $\widetilde{L}$ has dimensions of action. 
The Poincar\'e invariant Lagrangian in the parametric description \cite{InvariantLag} is the invariant and  homogeneous function of degree 1 in the derivatives of the boundary kinematical variables of both particles:
\[
\widetilde{L}_P=g\sqrt{\frac{\dot{x}_1^\mu\dot{x}_{2\mu}}{(x_1-x_2)^\mu(x_2-x_1)_\mu}}=g\sqrt{\frac{c^2\dot{t}_1\dot{t}_2-\dot{\bi r}_1\cdot\dot{\bi r}_2}{({\bi r}_1-{\bi r}_2)^2-c^2(t_1-t_2)^2}}.
\]
The variables ${\bi r}_a$ represent the location of the center of charge of the  corresponding Dirac particle. It is also invariant in the interchange $1\leftrightarrow2$ of the two particles.

The expression inside the squared root is dimensionless because the evolution parameter is dimensionless and, therefore, the coupling coefficient $g$ has dimensions of action. The interaction Lagrangian can also be rewritten as
\[
\widetilde{L}_P=g\frac{\sqrt{c^2-{\bi u}_1\cdot{\bi u}_2}}{\sqrt{({\bi r}_1-{\bi r}_2)^2-c^2(t_1-t_2)^2}}\sqrt{\dot{t}_1\dot{t}_2},
\]
where ${\bi u}_a=d{\bi r}_a/dt_a=\dot{\bi r}_a/{\dot{t}_a}$, is the velocity of the CC of the particle $a=1,2$. Please remark that to obtain a Poincar\'e invariant function we need to describe two different time coordinates, one for each particle. 

If we fix some particular inertial observer, the synchronous description in this frame produces the constraint $t_1=t_2=t$, and leads to:
\[
\widetilde{L}_P=g\frac{\sqrt{c^2-{\bi u}_1\cdot{\bi u}_2}}{|{\bi r}_1-{\bi r}_2|}\dot{t}.
\]
If the evolution parameter is chosen as the time
coordinate in this frame, $\dot{t}=1$, the homogeneity of the Lagrangian dissapears and also the  invariance under a Lorentz boost, because simultaneous events in one frame are not, in general, simultaneous in another. 
The Lagrangian in the synchronous time evolution description in any arbitrary inertial frame becomes:
\[
{L}_P=g\frac{\sqrt{c^2-{\bi u}_1\cdot{\bi u}_2}}{|{\bi r}_1-{\bi r}_2|},
\]
where now we have deleted the symbol $\;\widetilde{\;}$ over the Lagrangian which has now dimensions of energy.
In the limit of point particles is independent of the velocities and becomes the instantaneous Coulomb interaction. The constant $g$ is thus
\[
gc=-\frac{e^2}{4\pi\epsilon_0},\quad\Rightarrow\quad g=-\frac{e^2}{4\pi\epsilon_0c},
\]
and we get the above interaction Lagrangian (\ref{eq:LagrangianPoin}), where $u_1=u_2=c$, for Dirac particles. }
\end{quotation}

The dynamical equation corresponding to the particle 1 comes from
\[
\frac{\partial L_1}{\partial{\bi r}_1}-\frac{d}{dt}\left(\frac{\partial L_1}{\partial{\bi u}_1}\right)+
\frac{d^2}{dt^2}\left(\frac{\partial L_1}{\partial{\bi a}_1}\right)+\frac{\partial L_P}{\partial{\bi r}_1}
-\frac{d}{dt}\left(\frac{\partial L_P}{\partial{\bi u}_1}\right)=0,
\]
and the same for particle 2. Since the free Lagrangian $L_1$ is independent of the variables ${\bi r}_1$, the first three terms are reduced to
\[
-\frac{d}{dt}\left[\frac{\partial L_1}{\partial{\bi u}_1}-\frac{d}{dt}\left(\frac{\partial L_1}{\partial{\bi a}_1}\right)\right]=-\frac{d{\bi p}_1}{dt},
\]
where ${\bi p}_1=m\gamma(v_1){\bi v}_1$ is the mechanical linear momentum of particle 1, expressed in terms of the center of mass velocity of this particle.
If we take this term to the left hand side, we get
\[
\frac{d{\bi p}_1}{dt}=\frac{\partial L_P}{\partial{\bi r}_1}
-\frac{d}{dt}\left(\frac{\partial L_P}{\partial{\bi u}_1}\right)={\bi F}_1,
\]
where ${\bi F}_1$ is the total external force acting on particle 1. The expression of this force is:
\[
{\bi F}_1=\frac{e^2}{4\pi\epsilon_0c}\left[\frac{\sqrt{c^2-{\bi u}_1\cdot{\bi u}_2}}{|{\bi r}_1-{\bi r}_2|^3} \right] ({\bi r}_1-{\bi r}_2)
\]
\[
+\frac{e^2}{4\pi\epsilon_0c}\left[\frac{({\bi r}_1-{\bi r}_2)\cdot({\bi u}_1-{\bi u}_2)}{2|{\bi r}_1-{\bi r}_2|^3\sqrt{c^2-{\bi u}_1\cdot{\bi u}_2}}\right]{\bi u}_2
\]
\[
-\frac{e^2}{4\pi\epsilon_0c}\left[\frac{{\bi a}_1\cdot{\bi u}_2+{\bi u}_1\cdot{\bi a}_2}{4|{\bi r}_1-{\bi r}_2|(c^2-{\bi u}_1\cdot{\bi u}_2)^{3/2}}\right]{\bi u}_2
\]
\begin{equation}
-\frac{e^2}{4\pi\epsilon_0c}\left[\frac{1}{2|{\bi r}_1-{\bi r}_2|\sqrt{c^2-{\bi u}_1\cdot{\bi u}_2}} \right]{\bi a}_2
\label{PoincaForce1} 
\end{equation}
The first term is a repulsive force along the instantaneous relative separation between the CC's. If the velocity part ${\bi u}_1\cdot{\bi u}_2\approx0$, is negligible it represents the instantaneous Coulomb interaction between the charges. There are next two terms in the direction of the velocity of the CC of the other particle and finally another in the direction of the acceleration of the CC of particle 2. If the velocities are the same ${\bi u}_1\cdot{\bi u}_2=c^2$, the interaction Lagrangian vanishes. The force becomes singular. The term in the Lagrangian $\sqrt{c^2-{\bi u}_1\cdot{\bi u}_2}$, which arises by the relativistic modification of the Coulomb potential, depends on the relative orientation between the velocities of the CC's of the particles and can take a positive value in the range $[0,\sqrt{2}\,c]$. The highest value of this term can be obtained if the CC velocities are opposite to each other, as it happens in the situation which leads to the formation of bound pairs of Dirac particles which will be discussed later.

The accelerations that appear in some terms are the accelerations of the CC's of each particle. They satisfy the dynamical equation (\ref{eq:Could2rdt2})
\begin{equation}
{\bi a}_a=\frac{d^2{\bi r}_a}{dt^2}=\frac{c^2-{\bi u}_a\cdot{\bi v}_a}{({\bi q}_a-{\bi r}_a)^2}({\bi q}_a-{\bi r}_a), \quad a=1,2
\label{eq:CCdynam}
\end{equation}
in terms of the position ${\bi q}_a$ and velocity ${\bi v}_a$ of the CM and the position ${\bi r}_a$ and velocity ${\bi u}_a$ of the CC of the corresponding particle. In this way the term along ${\bi a}_2$ of formula (\ref{PoincaForce1}) is along the separation vector ${\bi q}_2-{\bi r}_2$.

The force on particle 2 is obtained from (\ref{PoincaForce1}) by replacing $1\leftrightarrow2$:
\[
{\bi F}_2=\frac{e^2}{4\pi\epsilon_0c}\left[\frac{\sqrt{c^2-{\bi u}_1\cdot{\bi u}_2}}{|{\bi r}_1-{\bi r}_2|^3} \right] ({\bi r}_2-{\bi r}_1)
\]
\[
+\frac{e^2}{4\pi\epsilon_0c}\left[\frac{({\bi r}_1-{\bi r}_2)\cdot({\bi u}_1-{\bi u}_2)}{2|{\bi r}_1-{\bi r}_2|^3\sqrt{c^2-{\bi u}_1\cdot{\bi u}_2}}\right]{\bi u}_1
\]
\[
-\frac{e^2}{4\pi\epsilon_0c}\left[\frac{{\bi a}_1\cdot{\bi u}_2+{\bi u}_1\cdot{\bi a}_2}{4|{\bi r}_1-{\bi r}_2|(c^2-{\bi u}_1\cdot{\bi u}_2)^{3/2}}\right]{\bi u}_1
\]
\begin{equation}
-\frac{e^2}{4\pi\epsilon_0c}\left[\frac{1}{2|{\bi r}_1-{\bi r}_2|\sqrt{c^2-{\bi u}_1\cdot{\bi u}_2}} \right]{\bi a}_1.
\label{PoincaForce2} 
\end{equation}

\subsection{The Poincar\'e force}
\label{PoinForce}

The instantaneous Coulomb force (\ref{CoulombForce}) is usually interpreted as the instantaneous electrostatic field of a point charge $e$ located at the point ${\bi r}_2$, evaluated at the point ${\bi r}_1$, times the charge $e$ located there. 

If we try to interpret the Poincar\'e force (\ref{PoincaForce1}) in terms of fields, the first term looks like the electrostatic Coulomb field of a point charge $e$, located at point ${\bi r}_2$, modulated by the velocity factor $\sqrt{c^2-{\bi u}_1\cdot{\bi u}_2}$, with a $1/r^2$ behavior. The second term, that depends on the velocity ${\bi u}_2$ with also a $1/r^2$ behavior, looks like the magnetic field of a point charge $e$ moving at the velocity ${\bi u}_2$ evaluated at ${\bi r}_1$ and depending also on the relative velocity ${\bi u}_1-{\bi u}_2$ of the charge in ${\bi r}_1$.
Finally, the other two terms that go like $1/r$ are acceleration terms, and can be interpreted as the radiation field of particle 2 acting instantaneously on particle 1.

Nevertheless, this interaction has been obtained without the mediation of an intermediate field. It is an action at a distance force that depends on the set of variables that define the kinematics of the two particles, at the same instant in the laboratory inertial reference frame, and by the requirement of Poincar\'e invariance of the Lagrangian. In this description there is no need to talk about the existence of interaction fields or even retarded fields of the particles interacting with each other.

\section{Conservation laws}
\label{conservation}

 The Lagrangians (\ref{eq:LagrangianCoul}) and (\ref{eq:LagrangianPoin}) are invariant under space and time translations and also under rotations, so that total energy $H_T$, total linear momentum ${\bi p}_T$ and total angular momentum with respect to the origin ${\bi J}_T$, are conserved observables. The Coulomb potential of (\ref{eq:LagrangianCoul}) is not invariant under Lorentz boosts and the interaction Lagrangian of (\ref{eq:LagrangianPoin}), although it is obtained by a Poincar\'e invariant analysis, is no longer invariant under Lorentz boosts because it has been restricted to a synchronous description in the laboratory frame, and simultaneity is not a conserved property in special relativity, under pure Lorentz transformations.

For the Poincar\'e invariant interaction, the total conserved energy is
\[
H_T=\gamma(v_1)mc^2+\gamma(v_2)mc^2+\frac{e^2}{4\pi\epsilon_0c}\frac{\sqrt{c^2-{\bi u}_1\cdot{\bi u}_2}}{|{\bi r}_1-{\bi r}_2|},
\]
the sum of the mechanical energy of each particle plus the instantaneous interaction energy.
 
The linear momentum of particle $a$ is the sum of the mechanical linear momentum ${\bi p}_a=m\gamma(v_a){\bi v}_a$, and the transfer of the linear momentum of the other particle, which is given by the term of the interaction Lagrangian ${\partial L_P}/{\partial{\bi u}_a}$.
For the particle 1 this linear momentum transfer ${\bi p}_{12}$, is
\[
{\bi p}_{12}=\frac{\partial L_P}{\partial{\bi u}_1}=\frac{e^2}{8\pi\epsilon_0c}\frac{{\bi u}_2}{|{\bi r}_1-{\bi r}_2|\sqrt{c^2-{\bi u}_1\cdot{\bi u}_2}},
\] 
while ${\bi p}_{21}\neq{\bi p}_{12}$, is
\[
{\bi p}_{21}=\frac{\partial L_P}{\partial{\bi u}_2}=\frac{e^2}{8\pi\epsilon_0c}\frac{{\bi u}_1}{|{\bi r}_1-{\bi r}_2|\sqrt{c^2-{\bi u}_1\cdot{\bi u}_2}},
\] 
of the same magnitude but of different direction because, in general, ${\bi u}_1\neq{\bi u}_2$.
What is conserved is the total linear momentum 
\begin{equation}
{\bi p}_T={\bi p}_1+{\bi p}_2+{\bi p}_{12}+{\bi p}_{21}.
\label{ConsPT}
\end{equation}

The conservation law of the total angular momentum with respect to the origin of any cartesian frame is
\[
{\bi J}_T={\bi r}_1\times({\bi p}_1+{\bi p}_{12})+{\bi r}_2\times({\bi p}_2+{\bi p}_{21})+{\bi S}_1+{\bi S}_2,
\]
where the spins ${\bi S}_a$, $a=1,2$, are the spins with respect to the CC's of each particle, as given in (\ref{eq:spinCC}).

In the case of the Coulomb interaction the total energy is conserved:
\[
H_T=\gamma(v_1)mc^2+\gamma(v_2)mc^2+\frac{e^2}{4\pi\epsilon_0c}\frac{1}{|{\bi r}_1-{\bi r}_2|}.
\]

For the linear momentum conservation, it is the sum of the mechanical linear momenta what is conserved,
\[
{\bi p}_T=\gamma(v_1)m{\bi v}_1+\gamma(v_2)m{\bi v}_2.
\]

The conservation of the total angular momentum reduces to the conservation of
\[
{\bi J}_T={\bi r}_1\times{\bi p}_1+{\bi r}_2\times{\bi p}_2+{\bi S}_1+{\bi S}_2.
\]  

We are going to analyze the electron-electron interaction from the point of view of the inertial laboratory observer. In this frame the synchronous description implies that the times of the two particles are the same and correspond to the time coordinate of the laboratory observer, $t_1=t_2=t$. The origin of this frame will be located at the center of mass of the two particles, so that the center of mass positions of both particles satisfy $\gamma(v_1){\bi q}_1(t)+\gamma(v_2){\bi q}_2(t)=0$, at any time. 

In addition to this requirement, the total linear momentum of the two particles, which is a conserved observable, will be zero. In the case of the Coulomb interaction this implies that ${\bi v}_1(t)=-{\bi v}_2(t)$ at any time for the velocities of the corresponding centers of mass.
The CM velocity of one of the particles is opposite to the CM velocity of the other and also
${\bi q}_1(t)=-{\bi q}_2(t)$.

In the case of the Poincar\'e invariant interaction this requirement does not hold because what is conserved is the total linear momentum ${\bi p}_T=0$. If the initial velocity of one of the particles is given, say ${\bi v}_1(0)$, the initial velocity of the other ${\bi v}_2(0)$ is restricted by this conservation of the total linear momentum and depends also on the particle separation and of the velocities of the CM and CC of both particles. This has to be taken into account when defining later the boundary conditions of the two-particle system to integrate the corresponding differential equations.

\section{Natural units}
\label{Natural}
The system of differential equations for both interactions (\ref{eq:Could2qdt2}) and (\ref{eq:Could2rdt2}) can be rewritten in terms of dimensionless variables. 
If we replace ${\bi u}=c\widetilde{\bi u}$,
${\bi v}=c\widetilde{\bi v}$, ${\bi r}=2R_0\widetilde{\bi r}$,  ${\bi q}=2R_0\widetilde{\bi q}$ and $t=(2R_0/c)\widetilde{t}$, where $R_0=\hbar/2mc$, is the separation between the CC and CM of the Dirac particle at rest
the equation (\ref{eq:d2rdt2}) becomes
\[
\frac{d^2{\bi r}}{dt^2}=\frac{c^2}{2R_0}\frac{d^2\widetilde{\bi r}}{d\widetilde{t}^2}=\frac{c^2}{2R_0}
\frac{1-{\widetilde{\bi v}}\cdot{\widetilde{\bi u}}}{({\widetilde{\bi q}}-{\widetilde{\bi r}})^2}({\widetilde{\bi q}}-{\widetilde{\bi r}}),
\]
the coefficients cancel and the equation remains of the same form as in (\ref{eq:d2rdt2}) in terms of the dimensionless variables and $c=1$.
The same thing happens to the other equation.
\[
m\frac{d\gamma(v){\bi v}}{dt}=\frac{mc^2}{2R_0}\frac{d\gamma(v)\widetilde{\bi v}}{d\tilde{t}}={\bi F}_1
\]
in terms of dimensionless variables and now $v<1$. 
The Coulomb force ${\bi F}_1$, (\ref{CoulombForce}) written in terms of natural units is
\[
{\bi F}_1=\frac{e^2}{4\pi\epsilon_04R_0^2}\frac{\widetilde{\bi r}_1-\widetilde{\bi r}_2}{|\widetilde{\bi r}_1-\widetilde{\bi r}_2|^3}.
\]
The numerical factor of the left hand side is $mc^2/2R_0$ when passed to the right hand side and using that $R_0=\hbar/2mc$, defines as the constant factor the fine structure constant:
\[
\frac{e^2}{4\pi\epsilon_0mc^22R_0}=\frac{e^2}{4\pi\epsilon_0c\hbar}=\alpha.
\]

For the Poincar\'e interaction, the first term of the force (\ref{PoincaForce1}) when reescaled is:
\[
\frac{e^2}{4\pi\epsilon_0 4R_0^2}\left[\frac{\sqrt{1-\widetilde{\bi u}_1\cdot\widetilde{\bi u}_2}}{|\widetilde{\bi r}_1-\widetilde{\bi r}_2|^3}(\widetilde{\bi r}_1-\widetilde{\bi r}_2) \right]
\]
and with the factor $mc^2/2R_0$ of the left hand side the final factor is again $\alpha$. The remaining three terms lead to the same factor $\alpha$, when expressed in terms of dimensionless variables.

The differential system to be solved in both interactions is
 \begin{eqnarray}
\ddot{\bi q}_a&=&\frac{1}{\gamma({\dot{\bi q}_a})}\left({\bi F}_a-\dot{\bi q}_a({\bi F}_a\cdot\dot{\bi q}_a)\right)\label{eq:qa2}\\
\ddot{\bi r}_a&=&\frac{1-\dot{\bi q}_a\cdot\dot{\bi r}_a}{({\bi q}_a-{\bi r}_a)^2}({\bi q}_a-{\bi r}_a),\quad a=1,2\label{eq:ra2}
 \end{eqnarray}
where each force ${\bi{F}_a}$, $a=1,2$ is given in the Coulomb interaction by
\begin{equation}
{\bi F}_1=\alpha\frac{{\bi r}_1-{\bi r}_2}{|{\bi r}_1-{\bi r}_2|^3}=-{\bi F}_2,
\label{eq:CoulombForce}
\end{equation}
and all variables are dimensionless in this natural system of units.

In the Poincar\'e invariant interaction, the force on particle 1, ${\bi F}_1$ is 
\[
{\bi F}_1=\alpha\left[\frac{\sqrt{1-{\bi u}_1\cdot{\bi u}_2}}{|{\bi r}_1-{\bi r}_2|^3}\right] ({\bi r}_1-{\bi r}_2) 
-\alpha\left[\frac{1}{2|{\bi r}_1-{\bi r}_2|\sqrt{1-{\bi u}_1\cdot{\bi u}_2}} \right]{\bi a}_2+
\]
\begin{equation}
+\alpha\left[\frac{({\bi r}_1-{\bi r}_2)\cdot({\bi u}_1-{\bi u}_2)}{2|{\bi r}_1-{\bi r}_2|^3\sqrt{1-{\bi u}_1\cdot{\bi u}_2}}\right]{\bi u}_2
-\alpha\left[\frac{{\bi a}_1\cdot{\bi u}_2+{\bi u}_1\cdot{\bi a}_2}{4|{\bi r}_1-{\bi r}_2|(1-{\bi u}_1\cdot{\bi u}_2)^{3/2}}\right]{\bi u}_2,\label{fuerzaPoincare1}
\end{equation}
and the force  on particle 2, ${\bi F}_2$ is obtained from this by replacing $1\leftrightarrow2$,  
\[
{\bi F}_2=\alpha\left[\frac{\sqrt{1-{\bi u}_1\cdot{\bi u}_2}}{|{\bi r}_1-{\bi r}_2|^3}\right] ({\bi r}_2-{\bi r}_1) 
-\alpha\left[\frac{1}{2|{\bi r}_1-{\bi r}_2|\sqrt{1-{\bi u}_1\cdot{\bi u}_2}} \right]{\bi a}_1+
\]
\begin{equation}
+\alpha\left[\frac{({\bi r}_1-{\bi r}_2)\cdot({\bi u}_1-{\bi u}_2)}{2|{\bi r}_1-{\bi r}_2|^3\sqrt{1-{\bi u}_1\cdot{\bi u}_2}}\right]{\bi u}_1
-\alpha\left[\frac{{\bi a}_1\cdot{\bi u}_2+{\bi u}_1\cdot{\bi a}_2}{4|{\bi r}_1-{\bi r}_2|(1-{\bi u}_1\cdot{\bi u}_2)^{3/2}}\right]{\bi u}_1,\label{fuerzaPoincare2}
\end{equation}
The difference with the Coulomb interaction is that the Poincar\'e invariant interaction does not satisfy Newton's third law, ${\bi F}_1\neq-{\bi F}_2$, the action-reaction principle.
All terms of equations (\ref{fuerzaPoincare1}-\ref{fuerzaPoincare2}) which depend on the acceleration of the charges
have been replaced by the expressions of (\ref{eq:ra2}). If the charges of the particles
are opposite, equation (\ref{eq:qa2}) will be affected by a change of sign.

These differential equations (\ref{eq:qa2}) and (\ref{eq:ra2}) are valid in any arbitrary inertial reference frame. The over dot represents the derivative with respect to the dimensionless time $\widetilde{t}$. In natural time units, the velocity of the CC is 1, the circular motion of the CC in the CM frame of the particle is of radius 1/2, the angular velocity is 2 and the time for a turn is of value $\pi$, while if the CM is moving with velocity $v$ this time for a turn is $\gamma(v)\pi$, and, in general, the separation between the CC and CM is not a constant of the motion, as we shall discuss in the next section. The only physical parameter that controls the interaction in the Coulomb and Poincar\'e invariant interaction is $\alpha$, the fine structure constant. 

In previous versions of the formalism we took the length scale $R_0=\hbar/2mc=1$. If we want to take as the natural unit of mass of the electron $m=1$, together with universal constants $\hbar=c=1$, this will imply that $R_0=1/2$ as we have done in this work. In this case the spin at the CM frame takes the form
${\bi S}=-m{\bi r}\times{\bi u}$ and with $|{\bi r}|=R_0=1/2$, $m=1$,  $|{\bi u}|=c=1$ we obtain that the natural unit of the spin $S=1/2$. The natural unit of charge is defined through the fine structure constant by redefining the natural unit of the permittivity of the vacuum $\epsilon_0$ as:
\[
\alpha=\frac{e^2}{4\pi\epsilon_0 c\hbar}=0.007297, \quad\rightarrow\quad e=1\;{\rm n.u.}\quad\Rightarrow \quad \frac{1}{4\pi\epsilon_0}=\alpha.
\]
What we have is to translate the international system of units to this natural system of units.
The relationship for the fundamental units of mass [M], length [L], time [T] and electric charge [Q] is:
\[
1\; {\rm n.u.} [M]=m_e=9.109534 \cdot10^{-31}{\rm Kg},\quad\rightarrow 1\;{\rm Kg}\equiv 1.09775\cdot10^{30}\;{\rm n.u.}
\]
\[
1\; {\rm n.u.} [L]=2R_0=\frac{\hbar}{mc}=3.86153\cdot10^{-13}{\rm m},\quad\rightarrow 1\; {\rm m}\equiv 2.58965\cdot10^{12}\;{\rm n.u.}
\]
\[
1\; {\rm n.u.} [T]=\tau_0=\frac{2R_0}{c}=6.44034\cdot10^{-22}\;{\rm s},\quad \rightarrow 1\; {\rm s}\equiv 7.76357\cdot10^{20}\;{\rm n.u.}
\]
\[
e=1 \;{\rm n.u.}\; [Q]=1.6021892\cdot10^{-19} {\rm C},\quad\rightarrow 1 \;{\rm C}\equiv 6.24146\cdot10^{18}\;{\rm n.u.}
\]
The electric field in the International System of units is expressed in V/m. According to the equivalence among units
\begin{equation}
1 \;{\rm V/m}=1\; {\rm m\, Kg\, s^{-2}\,C^{-1}},\quad 1\; {\rm V/m}=7.55676\cdot 10^{-19}\;{\rm n.u.} 
\label{voltM}
\end{equation}
The magnetic field is expressed in teslas. Since
\begin{equation}
1 \;{\rm T}=1\; {\rm Kg\, s^{-1}\,C^{-1}},\quad 1\; {\rm T}=2.26546\cdot 10^{-10}\;{\rm n.u.} 
\label{Teslanu}
\end{equation}
The rest mass energy of the electron $H_0=mc^2=1$ in natural units, in the international system of units is $H_0=8.18724\cdot10^{-14}\,$J. Then the unit of energy in the IS in terms of natural units is
\[
1\;{\rm n.u.}\;[E]\;H_0=mc^2=8.18724\cdot10^{-14}\,{\rm J},\qquad 1\;\; {\rm J}=1.22141\cdot10^{13}\;\;{\rm n.u.}
\]
If we express the energy in electron-Volts $H_0=mc^2/e=511003.37\,$eV. The conversion of eV into natural units is:
\[
1\;{\rm n.u.}\;[E]\;H_0=mc^2/e=511003.37\,{\rm eV},\qquad 1\;\;{\rm eV}=1.95693\cdot10^{-6}\;\;{\rm n.u.}
\]

\section{Boundary Conditions}

To integrate the system of differential equations (\ref{eq:qa2}) and (\ref{eq:ra2}), we have to establish the appropriate boundary conditions for the positions and velocities of both points, the CC and the CM of each particle.
These 12 boundary values for the variables ${\bi r}(0)$, ${\bi u}(0)$, ${\bi q}(0)$ and ${\bi v}(0)$ of each particle, are going to be expressed in terms of 10 parameters. This reduction is forced due to the 2 constraints: the absolute value of the CC velocity $|{\bi u}(0)|=1$, and that in the dynamical equations the vector ${\bi r}-{\bi q}$, has the direction of the acceleration of the CC and thus is orthogonal to the CC velocity ${\bi u}$. These 10 parameters are those parameters that define the relationship between the center of mass observer of the particle and any arbitrary inertial observer or laboratory observer who sees the CM of each particle moving at the speed ${\bi v}_a$. If the laboratory observer describes the two-particle system in their common center of mass frame, then the CM of the two particles is at rest and located at the origin of the reference frame. Since the Lagrangians (\ref{eq:LagrangianCoul}) and (\ref{eq:LagrangianPoin}) are invariant under space translations, the total linear momentum is conserved and the motion of the particles is such that, in the Coulomb interaction, 
${\bi q}_1(t)=-{\bi q}_2(t)$ and ${\bi v}_1(t)=-{\bi v}_2(t)$. In the Poincar\'e invariant interaction $\gamma(v_1){\bi q}_1(t)=-\gamma(v_2){\bi q}_2(t)$ holds but ${\bi p}_T=0$ does not imply that  ${\bi v}_1(t)=-{\bi v}_2(t)$. This requirement will be explicitely determined in the analysis of this interaction in the section {\bf\ref{Poincare}}.

If $t^*$ and ${\bi r}^*$ are the time and position of the CC of one of the particles for the center of mass observer $O^*$ of this particle, and $t$ and ${\bi r}$ are the time and position of the CC for any arbitrary inertial observer, they are related by the Poincar\'e transformation:
\[
x^{\mu}=\Lambda^\mu_\nu x^{*\nu}+a^\mu,\quad x^\mu\equiv(ct,{\bi r}),\quad x^{*\mu}\equiv(ct^*,{\bi r}^*), \quad a^\mu\equiv(cb,{\bi d}),
\]
and $\Lambda=L({\bi v})R(\psi,\theta,\phi)$ is a general Lorentz transformation, as a composition of a rotation $R$, followed by a boost of velocity ${\bi v}$, or pure Lorentz transformation $L({\bi v})$.

We are going to describe now the rotation $R(\psi,\theta,\phi)$, the boost or pure Lorentz transformation $L({\bi v})$ with arbitrary velocity ${\bi v}$, the space translation  $T({\bi d})$ of displacement ${\bi d}$, and finally the time translation $T(b)$, of value $b$. The velocity $v$ of absolute value $v$ and thenithal angle $\beta$ and azimuthal angle $\lambda$ is:
\begin{equation}
v_x=v\sin\beta \cos\lambda,\quad v_y=v\sin\beta \sin\lambda,\quad v_z=v\cos\beta.
\label{eq:velo}
\end{equation}

\begin{figure}[!hbtp]\centering%
\includegraphics[width=7cm]{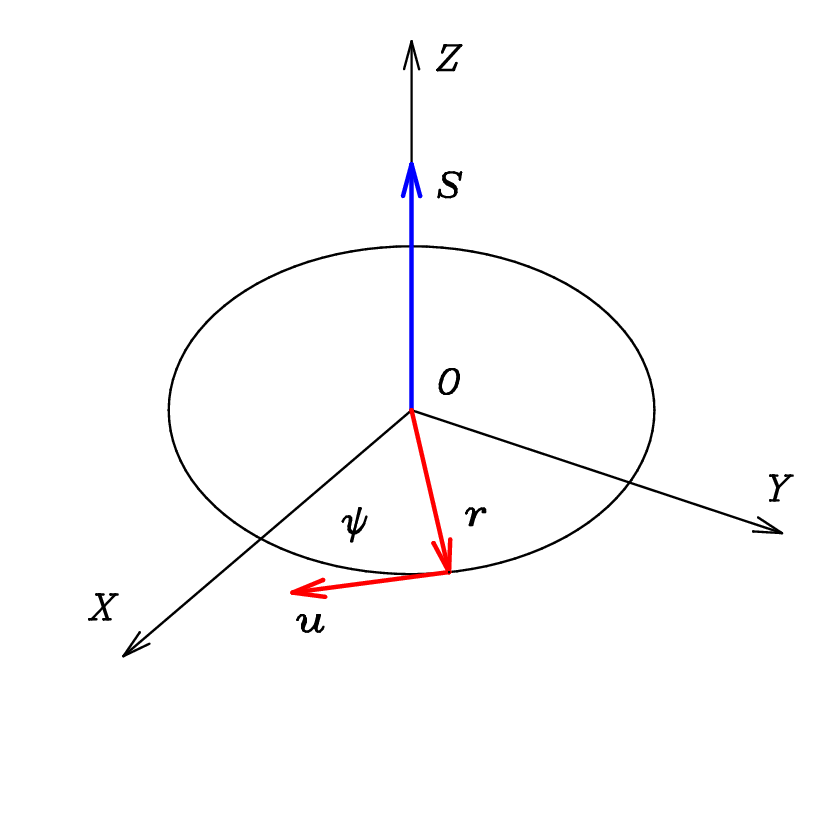}
\caption{Dirac particle in the center of mass reference frame, with the spin along the OZ axis. The initial position and velocity of the CC on the $XOY$ plane is determined by the phase  $\psi$. The radius of this motion is $R_0=1/2$, in natural units.} 
\label{fig2:initial}
\end{figure}

Let us first analyze the rotation. Let us assume we have the Dirac particle in the center of mass reference frame, as depicted in figure {\bf\ref{fig2:initial}}, with the spin along $OZ$ axis.

If we initially have the CC at the position $(R_0,0,0)$ on the $OX$ axis, we rotate first an angle $\psi$, around the $OZ$ axis to modify the CC position. Next, we change the orientation of the spin by modifying the zenithal angle $\theta$, and azimuthal angle $\phi$, so that the arbitrary rotation is the composition of the rotations
\[
R(\psi,\theta,\phi)=R_{OZ}(\phi)R_{OY}(\theta)R_{OZ}(\psi),
\]
\[
R_{OZ}(\psi)=\pmatrix{\cos\psi&-\sin\psi&0\cr \sin\psi&\cos\psi&0\cr 0&0&1},
\]
\[
R_{OY}(\theta)=\pmatrix{\cos\theta&0&\sin\theta\cr 0&1&0\cr -\sin\theta&0&\cos\theta },\quad R_{OZ}(\phi)=\pmatrix{\cos\phi&-\sin\phi&0\cr \sin\phi&\cos\phi&0\cr 0&0&1}.
\]
If initially at time $t^*_0$, the CC is at the point $(R_0,0,0)$, then the position, velocity and acceleration of the CC, before the rotation, are:
\[
{\bi r}^*(t^*_0)=R_0\pmatrix{1\cr 0\cr 0},\quad {\bi u}^*(t^*_0)=c\pmatrix{0\cr -1\cr 0}, \quad {\bi a}^*(t^*_0)=\frac{c^2}{R_0}\pmatrix{-1\cr 0\cr 0}.
\]
These variables, after the rotation, become ${\bi r}^*_0=R_{OZ}(\phi)R_{OY}(\theta)R_{OZ}(\psi){\bi r}^*(t^*_0)$, and the same for ${\bi u}^*(t^*_0)$ and ${\bi a}^*(t^*_0)$, at the same time $t^*_0$, and thus:
 \begin{equation}
{\bi r}^*_0=\frac{1}{2}\pmatrix{\cos\theta\cos\phi\cos\psi-\sin\phi\sin\psi\cr \cos\theta\sin\phi\cos\psi+\cos\phi\sin\psi\cr
-\sin\theta\cos\psi},
 \label{eq:valoresinicialesr}
 \end{equation}
  \begin{equation}
  {\bi u}^*_0=\pmatrix{\cos\theta\cos\phi\sin\psi+\sin\phi\cos\psi\cr \cos\theta\sin\phi\sin\psi-\cos\phi\cos\psi\cr
-\sin\theta\sin\psi},
 \label{eq:valoresinicialesu}
 \end{equation}
\begin{equation}
{\bi a}^*_0=-2\pmatrix{\cos\theta\cos\phi\cos\psi-\sin\phi\sin\psi\cr \cos\theta\sin\phi\cos\psi+\cos\phi\sin\psi\cr
-\sin\theta\cos\psi}=-4{\bi r}^*_0,
 \label{eq:valoresinicialesac}
\end{equation}
in natural units.

A Lorentz transformation of velocity ${\bi v}$, is given by the $4\times4$ matrix
 \begin{equation}
L({\bi v})=\pmatrix{\gamma&\gamma{v_x}& \gamma{v_y}& 
\gamma{v_z}\cr 
\gamma{v_x}&1+{\displaystyle v_x^2\gamma^2\over\displaystyle\gamma+1}&{\displaystyle v_xv_y\gamma^2\over\displaystyle\gamma+1}&{\displaystyle v_xv_z\gamma^2\over\displaystyle\gamma+1}\cr 
\gamma{v_y}&{\displaystyle v_yv_x\gamma^2\over\displaystyle\gamma+1}& 1+{\displaystyle v_y^2\gamma^2\over\displaystyle 
\gamma+1}&{\displaystyle v_yv_z\gamma^2\over\displaystyle\gamma+1}\cr 
\gamma{v_z}&{\displaystyle v_zv_x\gamma^2\over\displaystyle \gamma+1}&{\displaystyle v_zv_y\gamma^2\over\displaystyle \gamma+1}&1+{\displaystyle v_z^2\gamma^2\over\displaystyle \gamma+1}\cr},
 \label{eq:Tdev}
 \end{equation} 
in natural units and where $\gamma\equiv(1-v_x^2-v_y^2-v_z^2)^{-1/2}$. Finally the two translations $T({\bi d})$ and $T(b)$.
The corresponding time $t_0$ and position ${\bi r}_0$ of the CC of the same event, for any arbitrary inertial observer are:
\[
t_0=\gamma\left(t^*_0+{{\bi v}\cdot{\bi r}^*_0}\right)+b,\quad {\bi r}_0={\bi r}^*_0+\gamma{\bi v}t^*_0+\frac{\gamma^2}{1+\gamma}({\bi v}\cdot{\bi r}^*_0){\bi v}+{\bi d},
\]
where $b$ is the time translation, ${\bi d}$ is the space translation and ${\bi v}$ is the velocity of the center of mass observer $O^*$ as measured by the observer $O$. It represents, therefore, the velocity of the center of mass of the particle for the arbitrary laboratory observer $O$.

If the initial instant to integrate the equations in the reference frame of the Laboratory $O$, is the time $t_0=0$, this corresponds to
$\gamma t^*_0=-\gamma {\bi v}\cdot{\bi r}^*_0-b$, for the center of mass observer of the particle, and therefore the initial position of the CC for the laboratory observer at the initial instant  $t_0=0$, is:
\begin{equation}
{\bi r}_0={\bi r}^*_0-\frac{\gamma}{1+\gamma}({\bi v}\cdot{\bi r}^*_0){\bi v}-b{\bi v}+{\bi d}.
\label{ecuac:r0}
\end{equation}
For the remaning observables, 
 \begin{equation}
{\bi u}_0=\frac{{\bi u}^*_0+\gamma{\bi v}+\frac{\gamma^2}{(1+\gamma)}{({\bi v}\cdot{\bi u}^*_0){\bi v}}}{\gamma(1+{\bi v}\cdot{\bi u}^*_0)},
 \label{ecuac:u0}
 \end{equation}
 \begin{equation}
{\bi a}_0=\frac{(1+{\bi v}\cdot{\bi u}^*_0){\bi a}^*_0-({\bi v}\cdot{\bi a}^*_0){\bi u}^*_0-\frac{\gamma}{(1+\gamma)}({\bi v}\cdot{\bi a}^*_0){\bi v}}{\gamma^2(1+{\bi v}\cdot{\bi u}^*_0)^3},
 \label{ecuac:a0}
 \end{equation}
where ${\bi r}^*_0$, ${\bi u}^*_0$ and ${\bi a}^*_0$ are given above in (\ref{eq:valoresinicialesr}), (\ref{eq:valoresinicialesu}) and (\ref{eq:valoresinicialesac}), respectively.

With these values we have the initial boundary conditions for the variables ${\bi r}_0$, ${\bi u}_0$, ${\bi v}_0={\bi v}$. What we have left is ${\bi q}_0$, the initial position of the CM in this reference frame.

The definition of the center of mass position of any Dirac particle in any reference frame at any time is given in (\ref{defposCM})
\[
{\bi q}(t)={\bi r}(t)+\frac{1-{\bi v}(t)\cdot{\bi u}(t)}{{\bi a}(t)^2}\,{\bi a}(t),
\]
then the initial value of the CM ${\bi q}_0$, in the laboratory frame is:
\[
{\bi q}_0={\bi r}_0+\frac{1-{\bi v}\cdot{\bi u}_0}{{{\bi a}_0}^2}\,{\bi a}_0,
\]
in terms of the position, velocity and acceleration of the CC in this reference frame and of the velocity of the CM, all these variables defined at the initial time $t_0=0$.

As a summary, the boundary conditions at $t_0=0$, in the laboratory frame, are:
 \begin{equation}
{\bi r}_0={\bi r}^*_0-\frac{\gamma}{1+\gamma}({\bi v}\cdot{\bi r}^*_0){\bi v}-b{\bi v}+{\bi d}, 
 \label{eq:initialcond1}
 \end{equation}
 \begin{equation}
 {\bi u}_0=\frac{{\bi u}^*_0+\gamma{\bi v}+\frac{\gamma^2}{(1+\gamma)}{({\bi v}\cdot{\bi u}^*_0){\bi v}}}{\gamma(1+{\bi v}\cdot{\bi u}^*_0)},
 \label{eq:initialcond3}
 \end{equation}
 \begin{equation}
{\bi q}_0={\bi r}_0+\frac{1-{\bi v}\cdot{\bi u}_0}{{{\bi a}_0}^2}\,{\bi a}_0,
 \label{eq:initialcond2}
 \end{equation}
 \begin{equation} 
 {\bi v}_0={\bi v},
  \label{eq:initialcond4}
 \end{equation}
with ${\bi a}_0$ given in (\ref{ecuac:a0}),
and the magnitudes ${\bi r}^*_0$, ${\bi u}^*_0$ and ${\bi a}^*_0$, are those which appear in (\ref{eq:valoresinicialesr}-\ref{eq:valoresinicialesac}).

\begin{figure}[!hbtp]\centering%
\includegraphics[width=7cm]{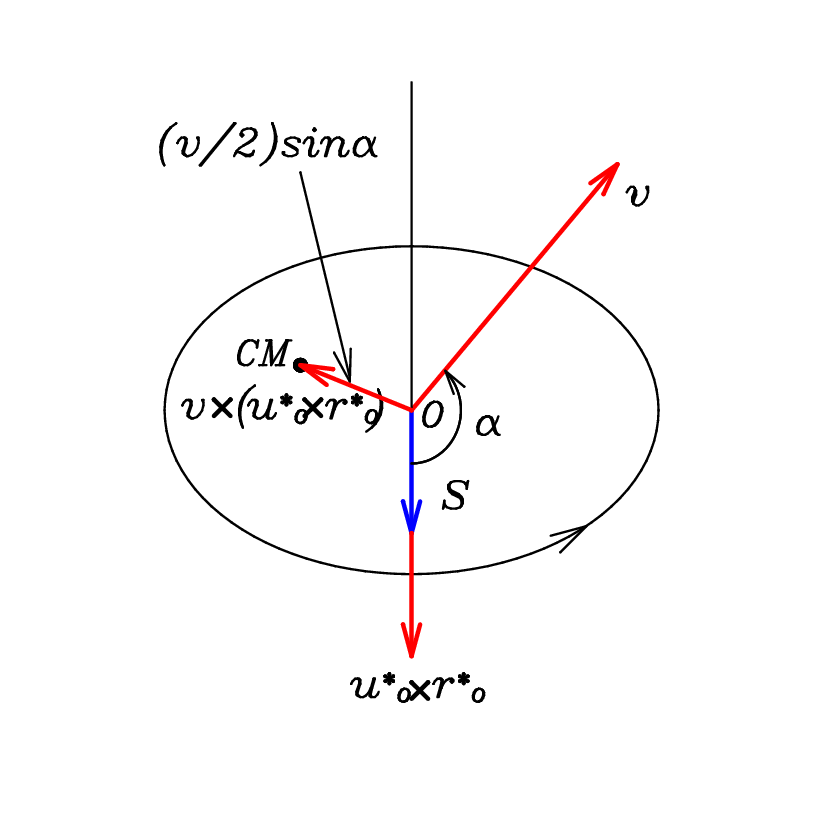}
\caption{Initial position of the CM when the velocity of the CM
${\bi v}$ and the spin direction form an angle $\alpha$. It is perpendicular to the vectors ${\bi v}$ and ${\bi u}^*_0\times{\bi r}^*_0$, of distance to the center $O-CM=(v/2)\sin\alpha$, in natural units, and is independent of the initial position and velocity of the CC. In general, the separation between both centers CC and CM, is in the interval $|{\bi r}-{\bi q}|\in[1/2-(v/2)\sin\alpha,1/2+(v/2)\sin\alpha]$, in natural units, so that this separation is not a constant of the motion.} 
\label{fig3:posCMini}
\end{figure}

Since ${\bi a}^*_0=-4{\bi r}^*_0$, taking the squared of (\ref{ecuac:a0}) we get
\[
{{\bi a}_0}^2=\frac{4}{\gamma^4(1+{\bi v}\cdot{\bi u}^*_0)^4},
\]
From (\ref{ecuac:u0}) the term
\[
1-{\bi v}\cdot{\bi u}_0=\frac{1}{\gamma^2(1+{\bi v}\cdot{\bi u}^*_0)},
\]
\[
\frac{1-{\bi v}\cdot{\bi u}_0}{{\bi a}_0^2}{\bi a}_0=\frac{\gamma^2}{4}(1+{\bi v}\cdot{\bi u}^*_0)^3{\bi a}_0,
\]
and if we substitute for ${\bi a}_0$ the expression (\ref{ecuac:a0}), where ${\bi a}^*_0=-4{\bi r}^*_0$, the initial position of the CM, ${\bi q}_0$, instead of (\ref{eq:initialcond2}), is
\begin{equation}
{\bi q}_0
={\bi v}\times({\bi u}^*_0\times{\bi r}^*_0)-b{\bi v}+{\bi d}.
\label{inicialcond4}
\end{equation}

In general, the initial position of the CM for the laboratory observer, given in (\ref{inicialcond4}), is contained in the zitterbewegung plane (with ${\bi d}=0=b$),
is independent of the initial position of the CC, ${\bi r}^*_0$,
and is depicted in the figure {\bf\ref{fig3:posCMini}}.
The distance to the center is $(v/2)\sin\alpha$.
If ${\bi v}$ is perpendicular to the ziterbewegung plane, ${\bi v}\cdot{\bi r}^*_0={\bi v}\cdot{\bi u}^*_0=0$, and the CM is at the center of the circle and becomes equidistant of the trajectory of the CC. If ${\bi v}$, has a different orientation, the separation between both centers CC and CM, $|{\bi r}-{\bi q}|$, is not a constant of the motion. It is in the range $|{\bi r}-{\bi q}|\in[R_0-(v/2)\sin\alpha, R_0+(v/2)\sin\alpha].$ 

We must remark that the CC ${\bi r}_0$ is the Poincar\'e transformed of the point ${\bi r}^*_0$, but 
${\bi q}_0$ is not the Poincar\'e transformed of the CM ${\bi q}^*_0$ in $O^*$. It is the CM of the particle in the laboratory frame $O$, at the same time $t_0$ than ${\bi r}_0$, while ${\bi r}^*_0$ and ${\bi q}^*_0$ are considered simultaneous at the frame $O^*$. Simultaneous events in one frame are not simultaneous in another. The point ${\bi q}_0$ represents the location of the CM of the particle simultaneous with ${\bi r}_0$, at the laboratory observer.

\section{Coulomb interaction boundary conditions}
\label{Coulomb}
We establish first the boundary conditions for the Coulomb interaction.

The 12 boundary conditions for particle 1 with a CM velocity ${\bi v}$, at the laboratory and in the natural system of units are:
 \begin{equation}
{\bi r}_{10}={\bi r}^*_{10}-\frac{\gamma}{1+\gamma}({\bi v}\cdot{\bi r}^*_{10}){\bi v}-\widetilde{b}_1{\bi v}+\widetilde{\bi d}_1,
 \label{eq:initialcond1RNatural}
 \end{equation}
  \begin{equation}
{\bi u}_{10}=\frac{{\bi u}^*_{10}+\gamma{\bi v}+\frac{\gamma^2}{(1+\gamma)}{({\bi v}\cdot{\bi u}^*_{10}){\bi v}}}
{\gamma(1+{\bi v}\cdot{\bi u}^*_{10})},
 \label{eq:initialcond1UNatural}
 \end{equation}
\begin{equation}
{\bi q}_{10}={\bi u}^*_{10}({\bi v}\cdot{\bi r}^*_{10})-{\bi r}^*_{10}({\bi v}\cdot{\bi u}^*_{10})-\widetilde{b}_1{\bi v}+\widetilde{\bi d}_1
,\quad {\bi v}_{10}={\bi v},
\label{inicialqyvNatural}
\end{equation}
where
${\bi r}^*_{10}$ and ${\bi u}^*_{10}$ are given, respectively in (\ref{eq:valoresinicialesr}) and (\ref{eq:valoresinicialesu}) for the angles $(\psi_1,\theta_1,\phi_1)$, and
\[
\widetilde{b}_1=(c/2R_0) b_1,\quad \widetilde{\bi d}_1={\bi d}_1/2R_0.
\]
The values of $b_1$ and ${\bi d}_1$ are in the international system of units and $\gamma\equiv\gamma({v})$.
For the particle 2, moving with velocity of its CM $-{\bi v}$, are
 \begin{equation}
{\bi r}_{20}={\bi r}^*_{20}-\frac{\gamma}{1+\gamma}({\bi v}\cdot{\bi r}^*_{20}){\bi v}+\widetilde{b}_2{\bi v}+\widetilde{\bi d}_2,
 \label{eq:initialcond2RNatural}
 \end{equation}
  \begin{equation}
{\bi u}_{20}=\frac{{\bi u}^*_{20}-\gamma{\bi v}+\frac{\gamma^2}{(1+\gamma)}{({\bi v}\cdot{\bi u}^*_{20}){\bi v}}}
{\gamma(1-{\bi v}\cdot{\bi u}^*_{20})},
 \label{eq:initialcond2UNatural}
 \end{equation}
where
the values of ${\bi r}^*_{20}$ and ${\bi u}^*_{20}$ are given, respectively in (\ref{eq:valoresinicialesr}) and (\ref{eq:valoresinicialesu}) for the angles $(\psi_2,\theta_2,\phi_2)$ and 
\[
\widetilde{b}_2=(c/2R_0) b_2,\quad \widetilde{\bi d}_2={\bi d}_2/2R_0.
\]

For the initial CM position and velocity of particle 2,
\begin{equation}
{\bi q}_{20}=-{\bi u}^*_{20}({\bi v}\cdot{\bi r}^*_{20})+{\bi r}^*_{20}({\bi v}\cdot{\bi u}^*_{20})+{b}_2{\bi v}+\widetilde{\bi d}_2
,\quad {\bi v}_{20}=-{\bi v},
\label{inicialqyv2Natural}
\end{equation}
Since the laboratory frame is the center of mass frame of the two particles, we have
\[
{\bi q}_{10}+{\bi q}_{20}=0,
\]
and this implies that the translation coefficients $b_a$ and ${\bi d}_a$, are not independent:
\[
{\bi q}_{10}-{\bi u}^*_{20}({\bi v}\cdot{\bi r}^*_{20})+{\bi r}^*_{20}({\bi v}\cdot{\bi u}^*_{20})+\widetilde{b}_2{\bi v}+\widetilde{\bi d}_2=0,
\]
and therefore the term $\widetilde{b}_2{\bi v}+\widetilde{\bi d}_2$ of equation (\ref{eq:initialcond2RNatural}) should be replaced by
\[
\widetilde{b}_2{\bi v}+\widetilde{\bi d}_2=-{\bi q}_{10}+{\bi u}^*_{20}({\bi v}\cdot{\bi r}^*_{20})-{\bi r}^*_{20}({\bi v}\cdot{\bi u}^*_{20}).
\]
In this way the boundary conditions for the CC of particle 2, instead of equation (\ref{eq:initialcond2RNatural}) should be
\begin{equation}
{\bi r}_{20}={\bi r}^*_{20}-\frac{\gamma}{1+\gamma}({\bi v}\cdot{\bi r}^*_{20}){\bi v}-{\bi q}_{10}+{\bi u}^*_{20}({\bi v}\cdot{\bi r}^*_{20})-{\bi r}^*_{20}({\bi v}\cdot{\bi u}^*_{20}).
\label{inicialqyv2NaturalFinal}
\end{equation}

The restriction that the laboratory observer is the CM observer of the 2-particle system, and that the description is synchronous for the 2 particles, restricts 4 of the space-time translation parameters and the CM velocity of the other particle and we shall select only those corresponding to particle 1, $b_1$, ${\bi d}_1$,
and we delete the subscripts for them. As a summary, the independent boundary parameters are:
\[
\left[\psi_1,\theta_1,\phi_1,v,\beta,\lambda,b,dx,dy,dz;\psi_2,\theta_2,\phi_2\right],
\] 
{\it i.e.}, the 10 parameters which define the initial state of particle 1 and only the spin orientation parameters and phase of particle 2.

\section{Poincar\'e invariant interaction boundary conditions}
\label{Poincare}

The boundary conditions in natural units for particle 1, moving its CM at the velocity ${\bi v}$, at the laboratory are, simmilarly as the Coulomb case:
 \begin{equation}
{\bi r}_{10}={\bi r}^*_{10}-\frac{\gamma(v)}{1+\gamma(v)}({\bi v}\cdot{\bi r}^*_{10}){\bi v}-\widetilde{b}_1{\bi v}+\widetilde{\bi d}_1,
 \label{eq:initialcond1RNaturalPoin}
 \end{equation}
 \begin{equation}
{\bi u}_{10}=\frac{{\bi u}^*_{10}+\gamma(v){\bi v}+\frac{\gamma(v)^2}{(1+\gamma(v))}{({\bi v}\cdot{\bi u}^*_{10}){\bi v}}}
{\gamma(v)(1+{\bi v}\cdot{\bi u}^*_{10})},
 \label{eq:initialcond1UNaturalPoin}
 \end{equation}
\begin{equation}
{\bi q}_{10}={\bi u}^*_{10}({\bi v}\cdot{\bi r}^*_{10})-{\bi r}^*_{10}({\bi v}\cdot{\bi u}^*_{10})-\widetilde{b}_1{\bi v}+\widetilde{\bi d}_1
,\quad {\bi v}_{10}={\bi v},
\label{inicialqyvNaturalPoin}
\end{equation}
where
${\bi r}^*_{10}$ and ${\bi u}^*_{10}$ are given, respectively in (\ref{eq:valoresinicialesr}) and (\ref{eq:valoresinicialesu}) for the angles $(\psi_1,\theta_1,\phi_1)$, and
\[
\widetilde{b}_1=(c/2R_0) b_1,\quad \widetilde{\bi d}_1={\bi d}_1/2R_0.
\]
The values of $b_1$ and ${\bi d}_1$ are in the international system of units and $\gamma(v)\equiv\gamma({v})$.

The particle 2 at the laboratory is not moving at velocity $-{\bi v}$ but at a velocity ${\bi v}_2$ related to ${\bi v}$ by the conservation of the total linear momentum (\ref{ConsPT}) in natural units.
In principle, for particle 2 the boundary conditions in natural units are:
 \begin{equation}
{\bi r}_{20}={\bi r}^*_{20}-\frac{\gamma(v_2)}{1+\gamma(v_2)}({\bi v}_2\cdot{\bi r}^*_{20}){\bi v}_2-\widetilde{b}_2{\bi v}_2+\widetilde{\bi d}_2,
 \label{eq:initialcond2RNaturalPoin}
 \end{equation}
 \begin{equation}
 {\bi u}_{20}=\frac{{\bi u}^*_{20}+\gamma(v_2){\bi v}_2+\frac{\gamma(v_2)^2}{(1+\gamma(v_2))}{({\bi v}_2\cdot{\bi u}^*_{20}){\bi v}_2}}
{\gamma(v_2)(1+{\bi v}_2\cdot{\bi u}^*_{20})},
 \label{eq:initialcond2UNaturalPoin}
 \end{equation}
that are functions of the unknown velocity ${\bi v}_2$.
The values of ${\bi r}^*_{20}$ and ${\bi u}^*_{20}$ are given, respectively in (\ref{eq:valoresinicialesr}) and (\ref{eq:valoresinicialesu}) for the angles $(\psi_2,\theta_2,\phi_2)$.

For the initial CM position and velocity of particle 2,
\begin{equation}
{\bi q}_{20}={\bi u}^*_{20}({\bi v}_2\cdot{\bi r}^*_{20})-{\bi r}^*_{20}({\bi v}_2\cdot{\bi u}^*_{20})-\widetilde{b}_2{\bi v}_2+{\bi d}_2
,\quad {\bi v}_{20}={\bi v}_2,
\label{inicialqyv2NaturalPoin}
\end{equation}
Since the laboratory frame is the center of mass frame of the two particles, we have
\begin{equation}
\gamma(v_{10}){\bi q}_{10}+\gamma(v_{20}){\bi q}_{20}=0,
\label{q1yq2}
\end{equation}
and this implies that the translation coefficients $b_a$ and ${\bi d}_a$, are not independent:
\[
\gamma(v){\bi q}_{10}+\gamma(v_2)\left[{\bi u}^*_{20}({\bi v}_2\cdot{\bi r}^*_{20})-{\bi r}^*_{20}({\bi v}_2\cdot{\bi u}^*_{20})-\widetilde{b}_2{\bi v}_2+\widetilde{\bi d}_2\right]=0,
\]
and therefore the term $-\widetilde{b}_2{\bi v}_2+\widetilde{\bi d}_2$ of equations (\ref{eq:initialcond2RNaturalPoin}) and (\ref{inicialqyv2NaturalPoin}) should be replaced by
\[
-\widetilde{b}_2{\bi v}_2+\widetilde{\bi d}_2=-\gamma(v){\bi q}_{10}/\gamma(v_2)-{\bi u}^*_{20}({\bi v}_2\cdot{\bi r}^*_{20})+{\bi r}^*_{20}({\bi v}_2\cdot{\bi u}^*_{20}).
\]
In this way the boundary conditions for the CC of particle 2, instead of equation (\ref{eq:initialcond2RNaturalPoin}) should be
\begin{equation}
{\bi r}_{20}={\bi r}^*_{20}-\frac{\gamma(v_2)}{1+\gamma(v_2)}({\bi v}_2\cdot{\bi r}^*_{20}){\bi v}_2-\frac{\gamma(v)}{\gamma(v_2)}{\bi q}_{10}-{\bi u}^*_{20}({\bi v}_2\cdot{\bi r}^*_{20})+{\bi r}^*_{20}({\bi v}_2\cdot{\bi u}^*_{20}).
\label{inicialR2NaturalFinalPoin}
\end{equation}
What is left is that the velocity ${\bi v}_2$ is not independent of ${\bi v}$ and of the other variables, because of the total linear momentum conservation.

The conservation of the total linear momentum implies in the laboratory frame that at any time the magnitude ${\bi p}_T=0$:
\[
m\gamma(v_1){\bi v}_1+m\gamma(v_2){\bi v}_2+\frac{e^2}{8\pi\epsilon_0c} \frac{{\bi u}_1+{\bi u}_2}{|{\bi r}_1-{\bi r}_2|\sqrt{c^2-{\bi u}_1\cdot{\bi u}_2}}=0.
\]
In natural units the above expression becomes:
\begin{equation}
\gamma({v}_1){\bi v}_1+\gamma({v}_2){\bi v}_2+\frac{\alpha({\bi u}_1+{\bi u}_2)}{2|{\bi r}_1-{\bi r}_2|\sqrt{1-{\bi u}_1\cdot{\bi u}_2}}=0.
\label{conservarP}
\end{equation}

We must take into account that the unknown velocity ${\bi v}_2$ is also contained in the expressions of ${\bi u}_{20}$ and ${\bi r}_{20}$, (\ref{eq:initialcond2UNaturalPoin}) and (\ref{inicialR2NaturalFinalPoin}) respectively. To integrate the dynamical equations we need to define properly the boundary conditions of both particles. If we fix that the initial CM velocity of particle 1 is ${\bi v}$, the expression (\ref{conservarP}) represents a numerical equation which when solved for the three components of the initial CM velocity ${\bi v}_2$ of particle 2, will allow us to complete the set of initial variables. Once this numerical system is solved, we have to introduce the values of ${\bi v}_2$ in (\ref{inicialR2NaturalFinalPoin}) to define the initial CC of particle 2, ${\bi r}_{20}$; in (\ref{eq:initialcond2UNaturalPoin}) to define the initial CC velocity of particle 2, ${\bi u}_{20}$; the initial CM position of particle 2, ${\bi q}_{20}=-{\bi q}_{10}\gamma(v)/\gamma(v_{2})$, and finally the initial CM velocity of particle 2, the calculated ${\bi v}_2$.

In this way, the complete set of values which determines the initial boundary conditions for the analysis of the electron-electron interaction with the Poincaré invariant Lagrangian is reduced, like in the Coulomb case, to the set of values:
\begin{equation}
\left[\psi_1,\theta_1,\phi_1,v,\beta,\lambda,b,dx,dy,dz;\psi_2,\theta_2,\phi_2\right],
\label{bothboundary}
\end{equation}
{\it i.e.}, the 10 parameters which define the initial state of particle 1 and only the initial spin orientation parameters and phase of particle 2.

\section{Poincar\'e interaction examples}
\label{PoincarInter}

\begin{figure}
\includegraphics[scale=0.30]{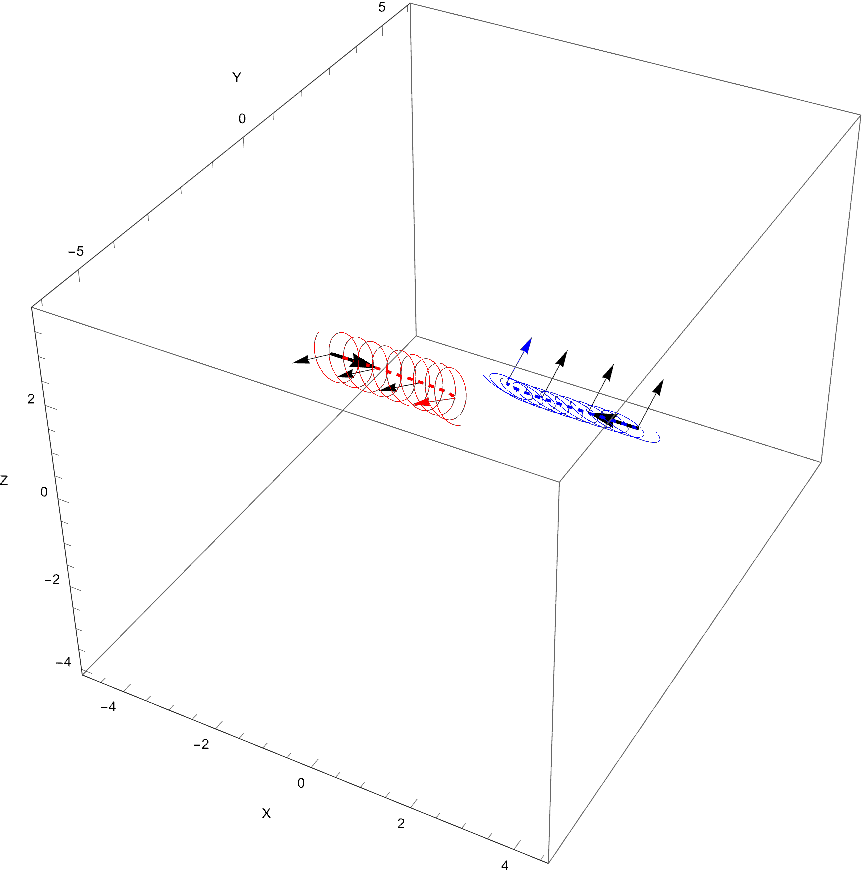}
\includegraphics[scale=0.30]{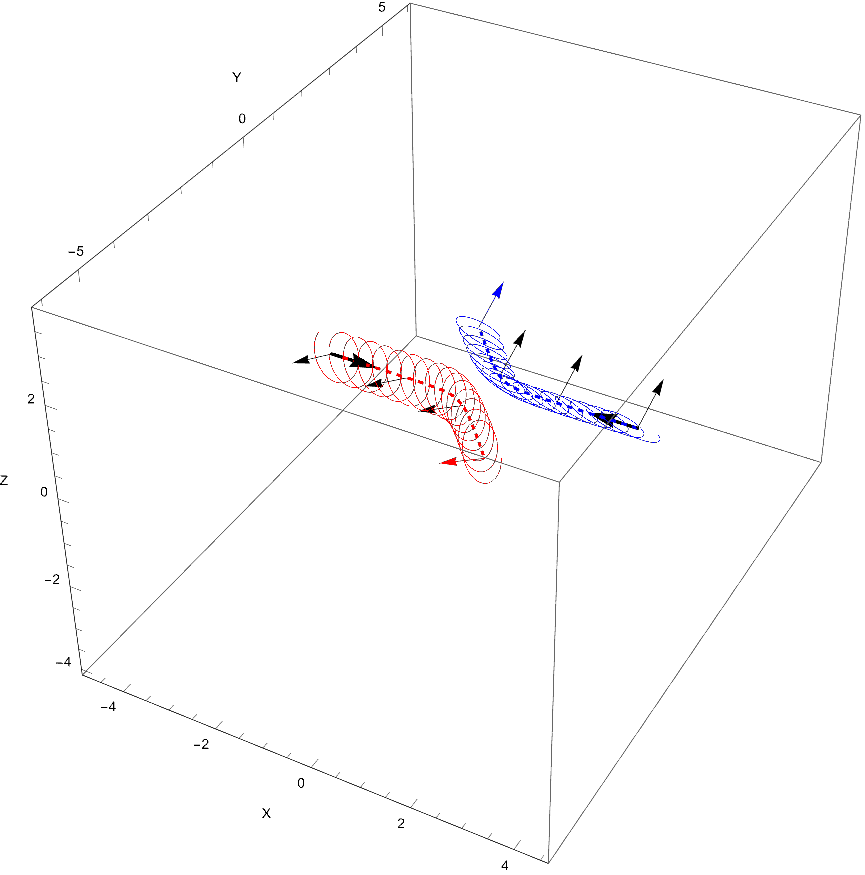}
\includegraphics[scale=0.30]{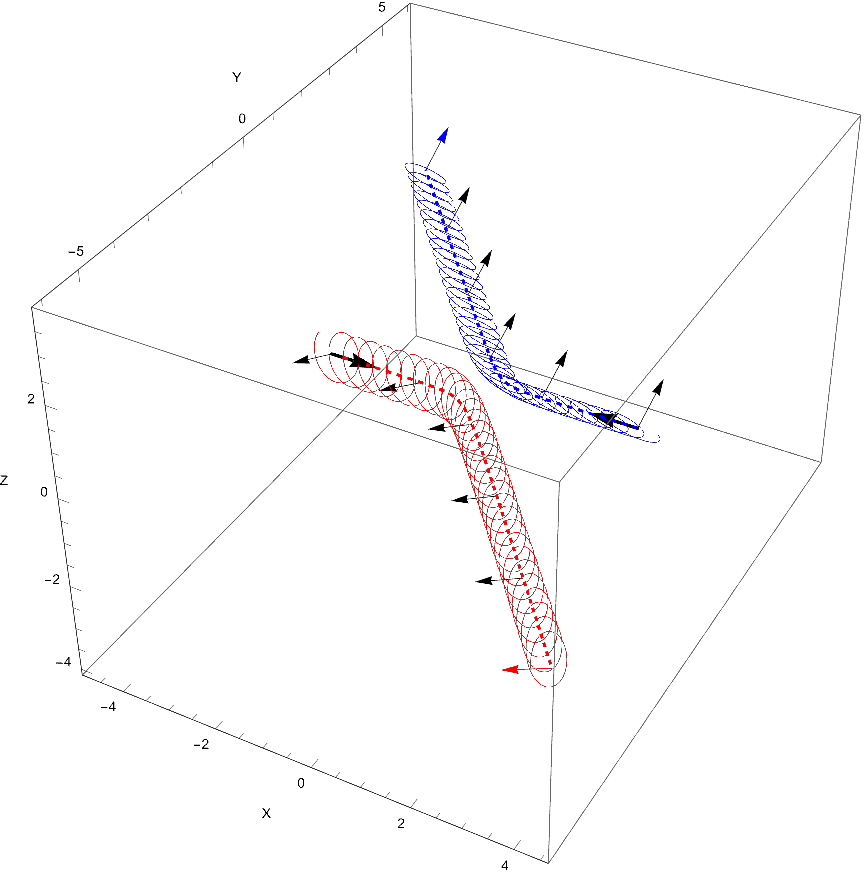}
\caption{Three instants of the Poincar\'e interaction of two Dirac particles with different spin orientations. It is depicted the initial velocity of both particles and in some places the CM spin to visualize the spin dynamics. It is analyzed in the center of mass of the two particles. The initial velocity is $v=0.08$, the initial location of particle 1 is $(3,0.3,0)$ and the spin orientations $\theta_1=30^\circ, \phi_1=60^\circ, \psi_1=0$ and $\theta_2=120^\circ,  \phi_2=180^\circ, \psi_2=180^\circ$. We see that the motion is almost free before and after the interaction section that is a region of a few Compton's wavelentgh. The spin $1/2$ has been reescaled and we see a slight change of orientation of both spins during the interaction.
}
\label{fig:Poincare}
\end{figure}

We are going to analyze different examples of the electron-electron interaction by using the Poincar\'e invariant interaction described by the Lagrangian (\ref{eq:LagrangianPoin}) and the Coulomb interaction Lagrangian (\ref{eq:LagrangianCoul}). The numerical computation is made with a Mathematica notebook \cite{Poincare}. In the next section we shall analyze the interaction by using both potentials \cite{both}. Readers interested can also analyze both interactions separately with the notebooks \cite{Coulomb} and. We have included in this notebook both integrations simoultaneously, the Coulomb and Poincar\'e invariant analysis, to compare the results under the same boundary conditions, for these two interaction potentials. In the figure {\bf\ref{fig:Poincare}}, with the help of the notebook \cite{Poincare}, we analyze the Poincar\'e interaction during three different integration times to see the evolution of the two particles.
It is depicted particle 1 in blue and particle 2 in red.  Full lines are used for the evolution of the CC's while dotted lines represent the motion of the CM's. It is also depicted the initial CM velocity of each particle and also at different CM positions the CM spin vector of each particle. These CM points where the spins are depicted are separated by the same integration time. When the integration stops we also depict at the final CM position of each particle their final CM spin, red for particle 1 and blue for particle 2, to see how the spin orientation has changed during the interaction. The spin $1/2$ has been reescaled because in natural units this vector will appear very small. 

\begin{figure}
\includegraphics[scale=0.35]{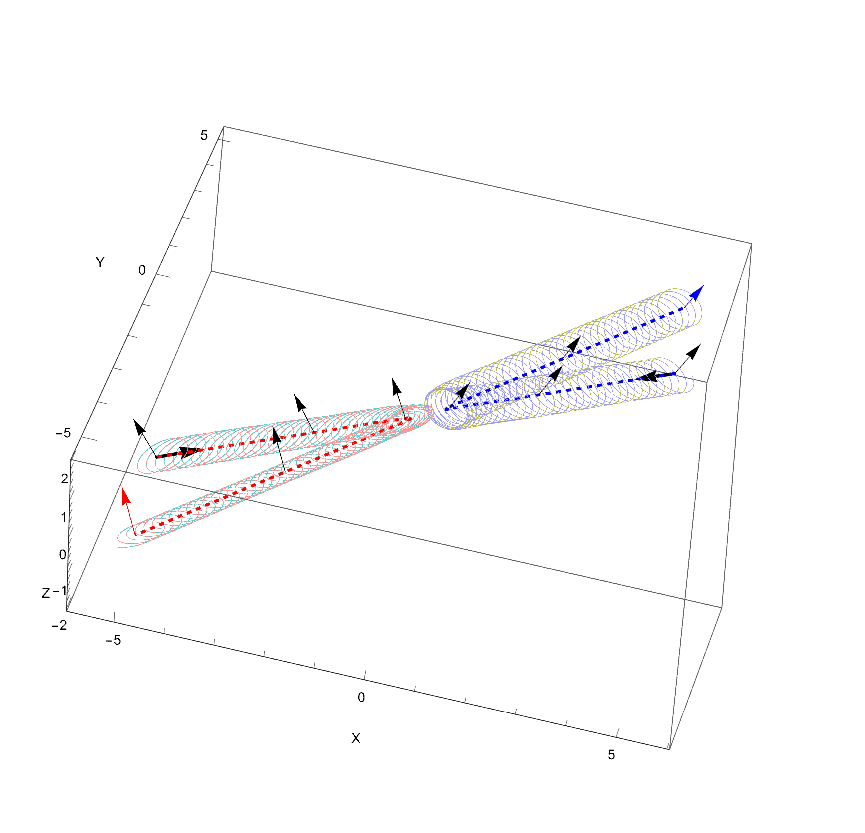}
\includegraphics[scale=0.35]{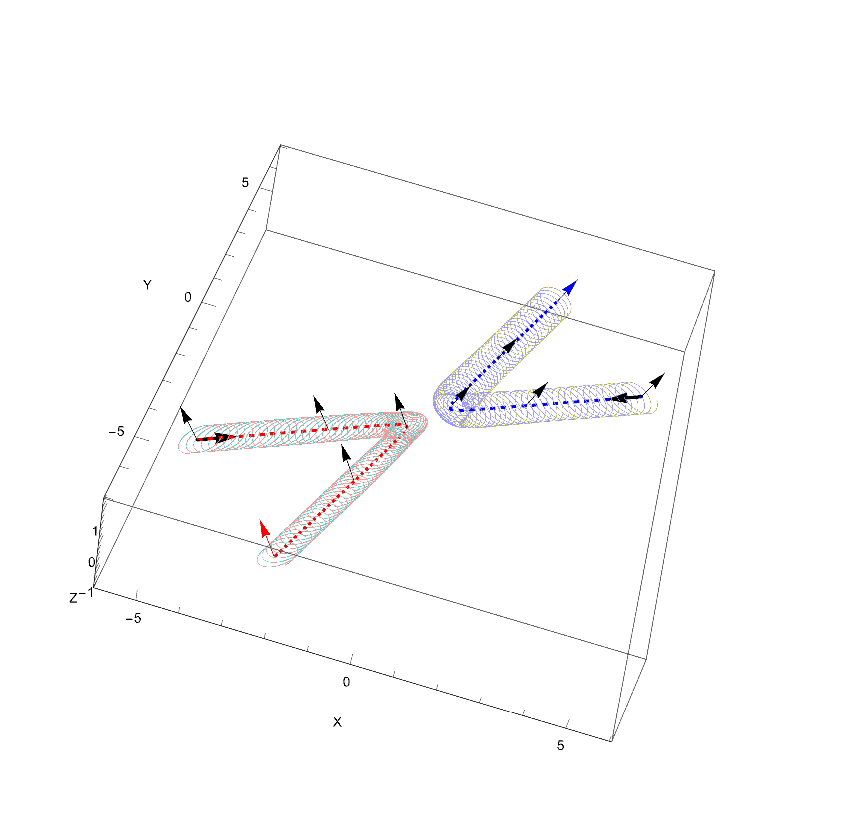}
\includegraphics[scale=0.35]{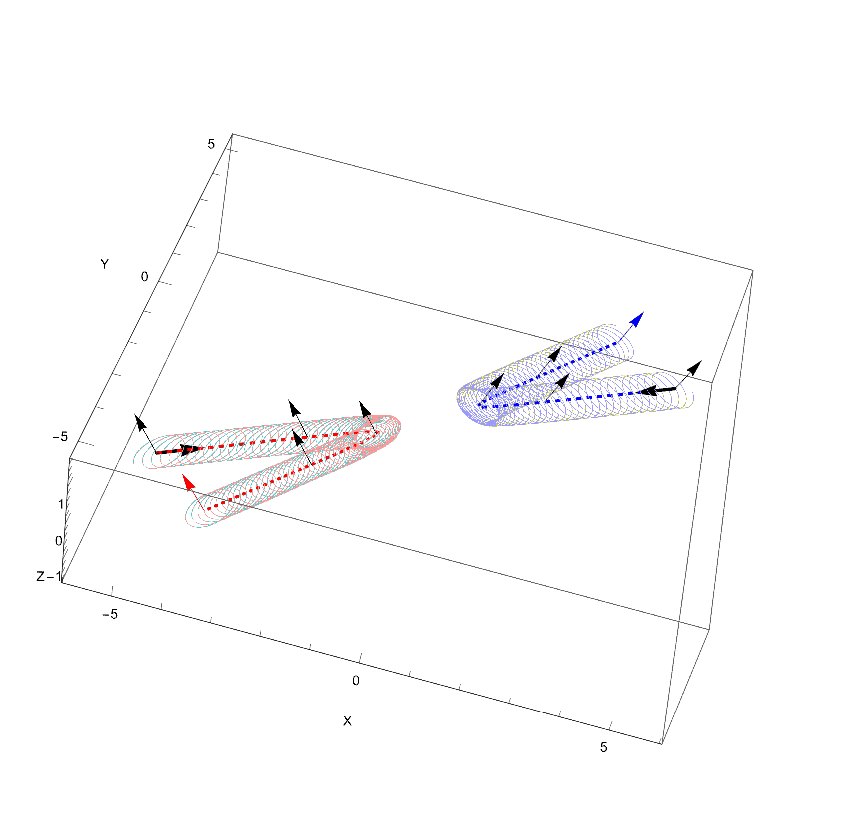}
\caption{Poincar\'e invariant interaction of two Dirac particles with boundary conditions in natural units $v=0.058$,
$\beta=90^\circ$, $\lambda=210^\circ$, initial position of the CM of particle 1, $dx=5,dy=3,dz=0$, spin orientation 
$\theta_1=20^\circ,\phi_1=0^\circ,\psi_1=30^\circ$ and for particle 2, $\theta_2=30^\circ,\phi_2=180^\circ,\psi_2=180^\circ$. The second and third pictures correspond only to the change of $\psi_2=90^\circ$, and $\psi_2=30^\circ$, respectively. The spin orientation of both particles has changed during the interaction.}
\label{repulsion}
\end{figure}
\newpage
In the figure {\bf\ref{repulsion}} we analyze three different repulsive interactions with the same boundary conditions except that the first the value of the CC phase of particle 2 (red) is $\psi_2=180^\circ$ and the next two take the values $\psi_2=90^\circ$ and $\psi_2=30^\circ$, respectively. In the first picture the CM's of the two particles arrive to a very close position while the simple change of the CC phase of particle 2 the repulsion occurs at a larger distance.

\section{Coulomb and Poincar\'e interaction examples}
\label{CoulombPoincarInter}
\begin{figure}
\includegraphics[scale=0.50]{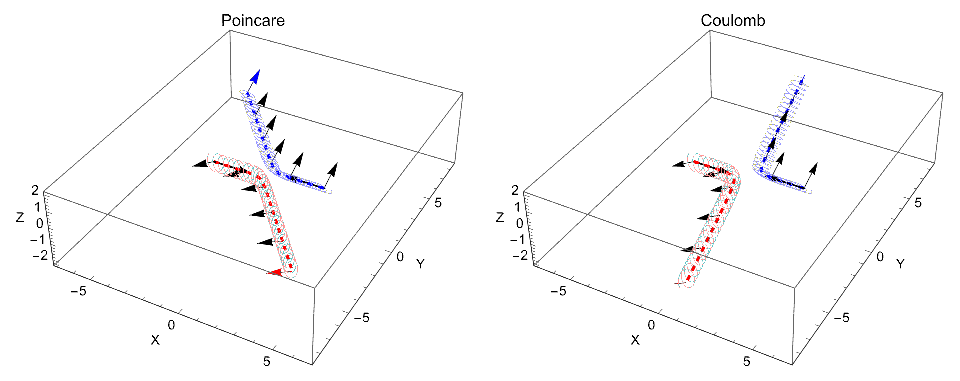}
\includegraphics[scale=0.50]{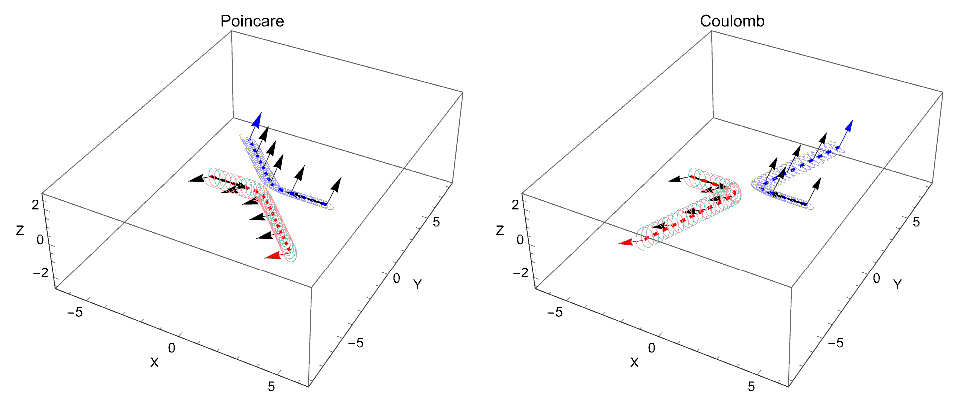}
\includegraphics[scale=0.50]{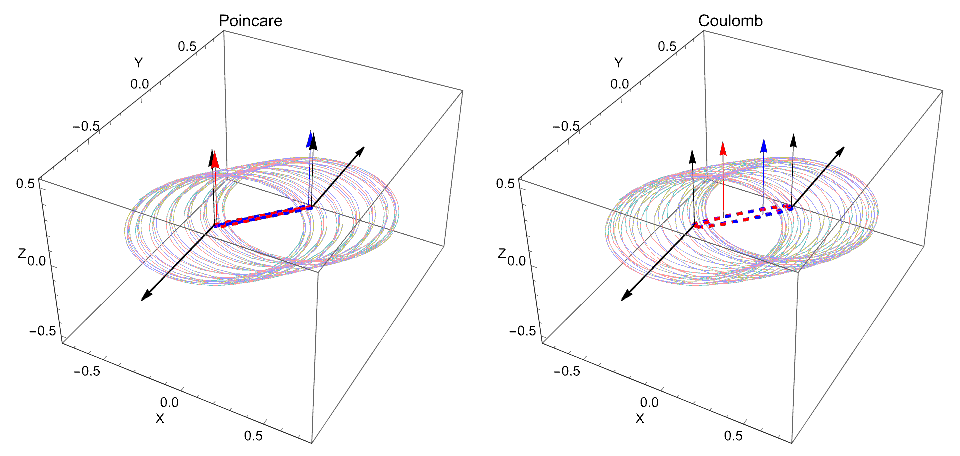}
\includegraphics[scale=0.50]{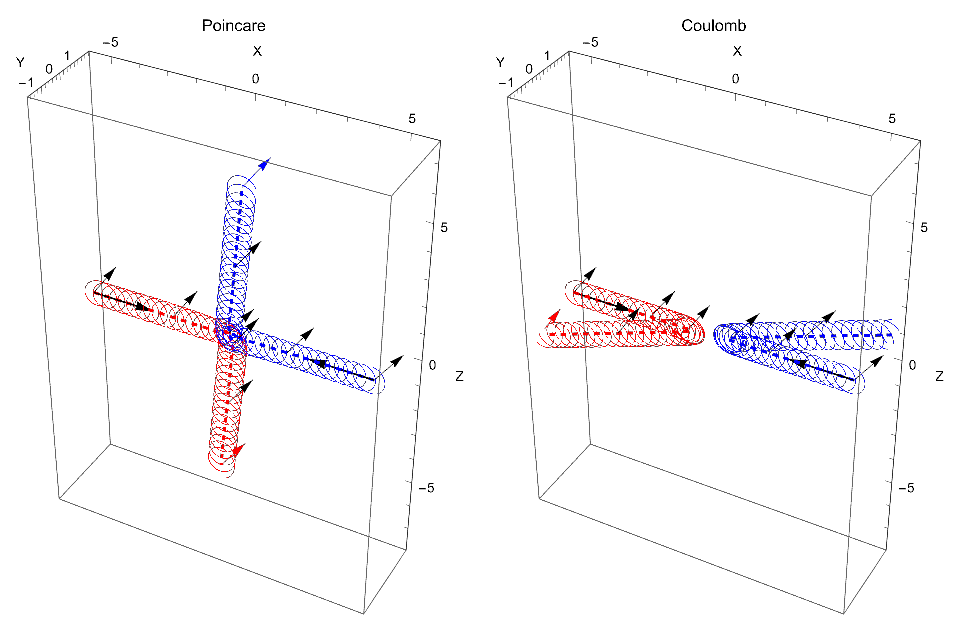}
\includegraphics[scale=0.50]{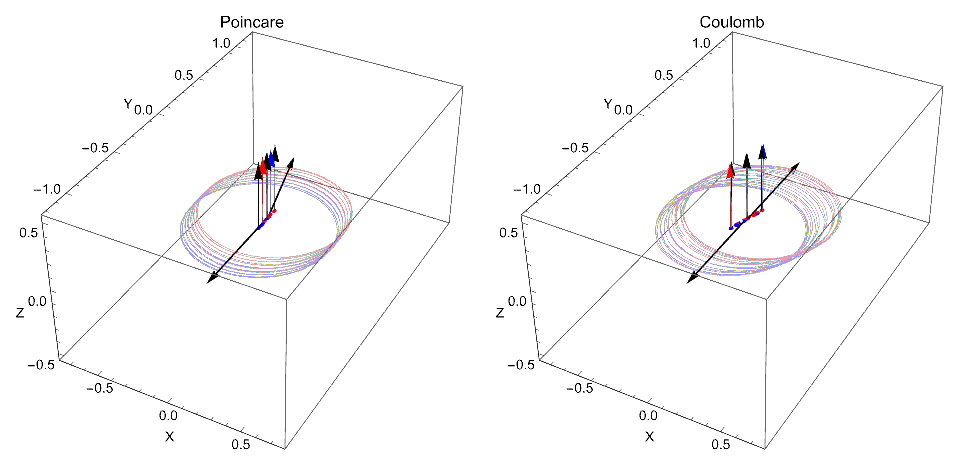}
\includegraphics[scale=0.55]{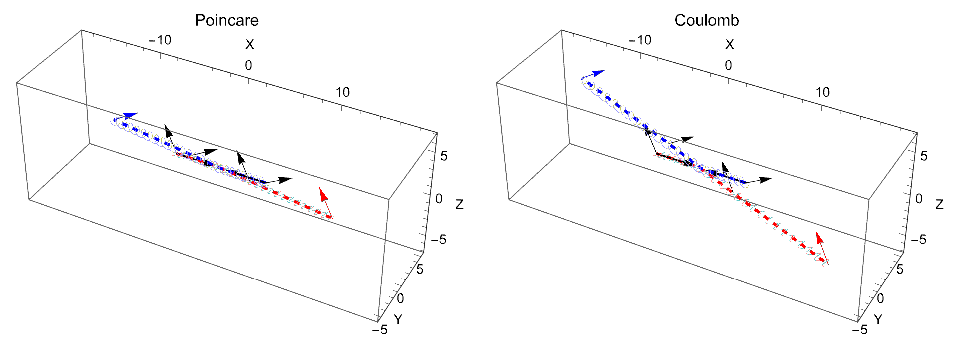}
\caption{Different interactions of two Dirac particles under the Poincar\'e invariant interaction of Lagrangian (\ref{eq:LagrangianPoin}) and also the Coulomb interaction of Lagrangian (\ref{eq:LagrangianCoul}) under the same boundary conditions of the two particles (\ref{bothboundary}). We show the boundary conditions different from zero.
From left to right and up and down:\\
{\tiny 1. $v=0.1,\beta=90^\circ,\lambda=180^\circ,dx=3,dy=0.3,\psi_2=180^\circ$.\\
2. The same values as in 1 except that $\psi_1=90^\circ,\psi_2=120^\circ$\\
3. $v=0.001, \beta=90^\circ,\lambda=90^\circ$, $\psi_1=0, \psi_2=180^\circ$.\\
4. $v=0.1, \beta=90^\circ,\lambda=90^\circ$, $\theta_1=35^\circ,\psi_1=120$, $\theta_2=35^\circ,\psi_1=50$, $dx=5$.\\
5. $v=0.01,\beta=90^\circ,\lambda=270^\circ,dx=0.001,\psi_1=0^\circ,\psi_2=120^\circ$\\
6. $\theta_1=60^\circ,\psi_1=29^\circ,v=0.2,\beta=90^\circ,\lambda=180^\circ,dx=5,dy=-0.1,\theta_2=45^\circ,\phi_2=150^\circ,\psi_2=50^\circ$.}
}
\label{fig:PoinCoul}
\end{figure}

We depict in the figure {\bf\ref{fig:PoinCoul}} Poincar\'e and Coulomb interactions, with the same
boundary conditions to appreciate the difference between both interaction potentials.
\begin{enumerate}
\item{Picture 1 represents the same Poincar\'e interaction as the one of the figure {\bf\ref{fig:Poincare}}. The two particles repel each other but the Coulomb interaction is different.}
\item{Picture 2 represents the same interaction as in the previous figure but the only change are the phases $\psi_1=90^\circ$ and $\psi_2=120^\circ$.}
\item{In picture 3 we show the formation of a bound pair of Dirac particles with parallel spins and apposite phases of the CC, $\psi-\psi_2=180^\circ$.}
\item{Picture 4 represents a right angle scattering under the Poincar\'e interaction while there is a repulsion for the Coulomb one.}
\item{Experiment 5 produces a another bound pair with parallel spins but the phase difference $\psi-\psi_2=120^\circ$. Remark that in the Coulomb interaction the initial velocities are always opposite to each other while in the Poincar\'e interaction are not because of total linear momentum conservation.}
\item{Experiment 6, with a velocity of $v=0.2$, this frontal interaction produces a forward scattering of the two particles, which cross to each other without modifying the motion of their CM's while the Coulomb interaction produces a deep scattering.}
\end{enumerate}

In all the mentioned examples, once the notebook is executed and activated the automatic execution mode, you can click on any of the boundary variables in the left hand pannel to modify the problem variables at your will, and a new integration is produced. You can see how small changes in the spin orientations or in the initial phases of the CC's, produce different scattering processes.
The dynamical equations are non-linear and very small changes in the initial conditions affect the outcome of the numerical experiment.
In the case of formation of bound pairs of electrons, the phases $\psi_1$ and $\psi_2$ do not need to be opposite to each other and you can determine which phase difference destroys the pairing. You can modify the CM velocity $v$ to reach the value of this velocity where the bound state is no longer stable. An external magnetic field along the spin direction does not destroy this bound state. 

The Coulomb interaction becomes singular when ${\bi r}_1={\bi r}_2$. The Poincar\'e invariant interaction becomes also singular when ${\bi u}_1={\bi u}_2$, even if the two particles are far away. The numerical integration breaks down when one of these two things happens and sometimes, in the Poincar\'e numerical integration, if the numerical value of the expression ${\bi u}_1\cdot{\bi u}_2>1$, the squared root $\sqrt{1-{\bi u}_1\cdot{\bi u}_2}$, is a complex number and the numerical integration fails. A small change in the boundary conditions can help to repare this fault, related to the numerical integration routine and not to the physical system. 

\begin{figure}
\includegraphics[scale=0.8]{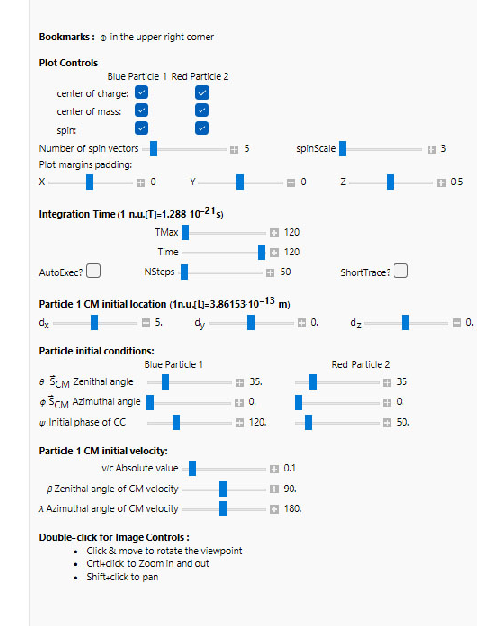}\includegraphics[scale=0.8]{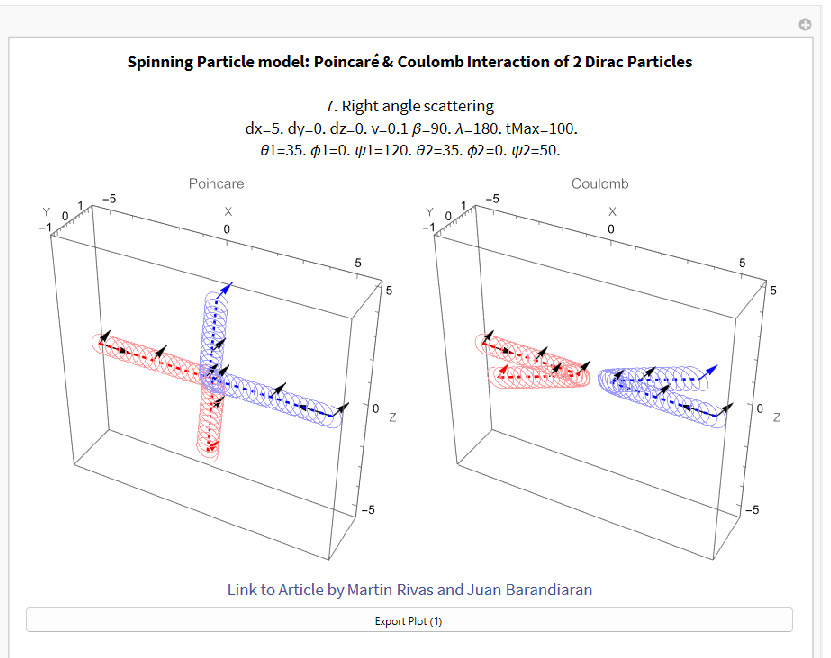}
\caption{The left figure represents the boundary variables integration panel. From top to bottom we find the 6 blue buttons that when activated the corresponding variables, the center of charge, center of mass and spin of both particles are depicted. The number of CM spin vectors to draw along the CM trajectory and the change of scale of the spin vectors.
You can change the maximum integration time, the time of integration and on the left the AutoExec button that, when activated, produces the integration step by step by dividing the total integration time in the number of selected steps.
In this way we see a film like description of the interaction of the Dirac particles. We end with the location of particle 1 and the characteristics of the spin orientation of both particles and the initial velocity of particle 1.
All these numbers can be changed with the + button or just by writting the corresponding new values on that variable by clicking on the old number.\\
The right figure is the screen where the integration is depicted. In the upper right corner, by pressing the + symbol, a new window is displayed where different bookmarks when activated, produce different integrations. Most of these correspond to the figures of this article. In the top of the figure it is also written the boundary variables of the corresponding integration and by pressing the {\tt Export Plot(1)} at the bottom, a figure in format .png is saved in the Notebook directory.}
\label{Panel}
\end{figure}

When Mathematica performs the integration the solution is depicted in a figure that has on the left hand side the boundary conditions panel, like the one depicted in the figure {\bf\ref{Panel}}. Interested readers can modify all of the boundary variables as they wish. In the figure caption of the figure  {\bf\ref{Panel}} it is explained how to modify the boundary variables and some other features. The Autoxec button produces consecutive integrations by dividing the integration time in short steps, showing a kind of a film of the scattering process. 

\section{Analysis of the Dirac particles interaction}
\label{final}
The numerical experiments in sections {\bf\ref{PoincarInter}} and {\bf\ref{CoulombPoincarInter}} are a few examples of the possibility of analyzing the interaction of two Dirac particles. At first glance there does not seem to be much difference between the Coulomb and the Poincar\'e invariant interaction before arriving to the interaction area. It seems that the extra terms of the Poincar\'e forces compared to the Coulomb force, produce a small local disturbance. But the velocity terms of the Poincar\'e Lagrangian produce a different outcome of the numerical experiment.

We are analyzing the interaction between two Dirac particles at a very small impact parameter of order of $10^{-13}$m, and very small separation between the CM's, which means that in some circunstances, the separation between the CC's is much smaller than this, and therefore, a very high interacting force. For a larger impact parameter we will be unable to distinguish between the CC and CM motions of the particles.

In all the examples we see that the initial motion of the particles is basically the same in both interactions, before reaching the interaction region. This region is a few times the internal radius $R_0$, and the motion of particles is almost free outside this region. The loops of the zitterbewegung motion are equally spaced outside this region, which means that the CM velocity is basically constant, except in the interaction area. The usual assumption in scattering theory that the initial and final states of a scattering experiment are free states can be confirmed, even at a very short distance of the scattering center, in these simulations. The total integration time is divided in equal steps and in every CM position of each particle it is depicted the corresponding CM spin. These number of spins drawn can be controlled with the {\tt Number of spin vectors} button, and the separation of these spins is a visual estimate of the CM velocity of the particles. 

We are going to compare these integrations with the analysis of the point particle electron-electron interaction under the instantaneous Coulomb interaction. Since for the spinless electron the CC and CM are the same point we locate each point electron at the same initial position of the CM of the corresponding Dirac particle. The initial velocity of each point particle is the same as the initial CM velocity of the corresponding Dirac particle, and we use the same impact parameter to describe these interactions. Because we have been unable to find a Poincar\'e invariant interaction for point particles, we shall restrict this analysis to the Coulomb interaction of point particles and also the Coulomb interaction of Dirac particles. The numerical integration is performed with the Mathematica notebook
\cite{Coulomb}. 

\begin{figure}
\includegraphics[scale=0.3]{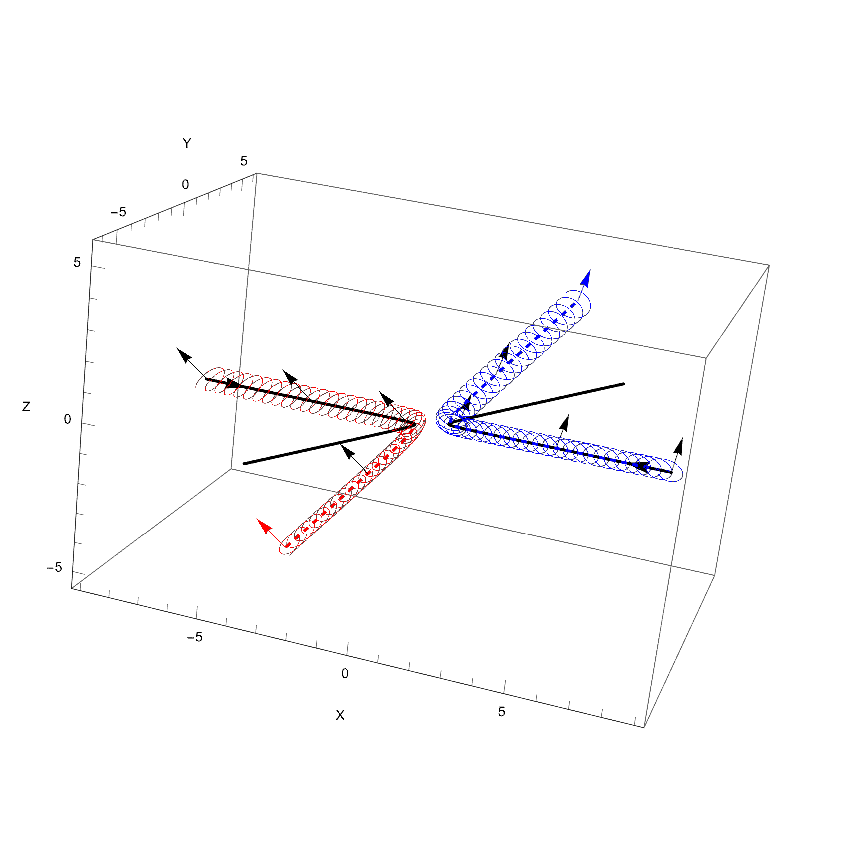}
\includegraphics[scale=0.35]{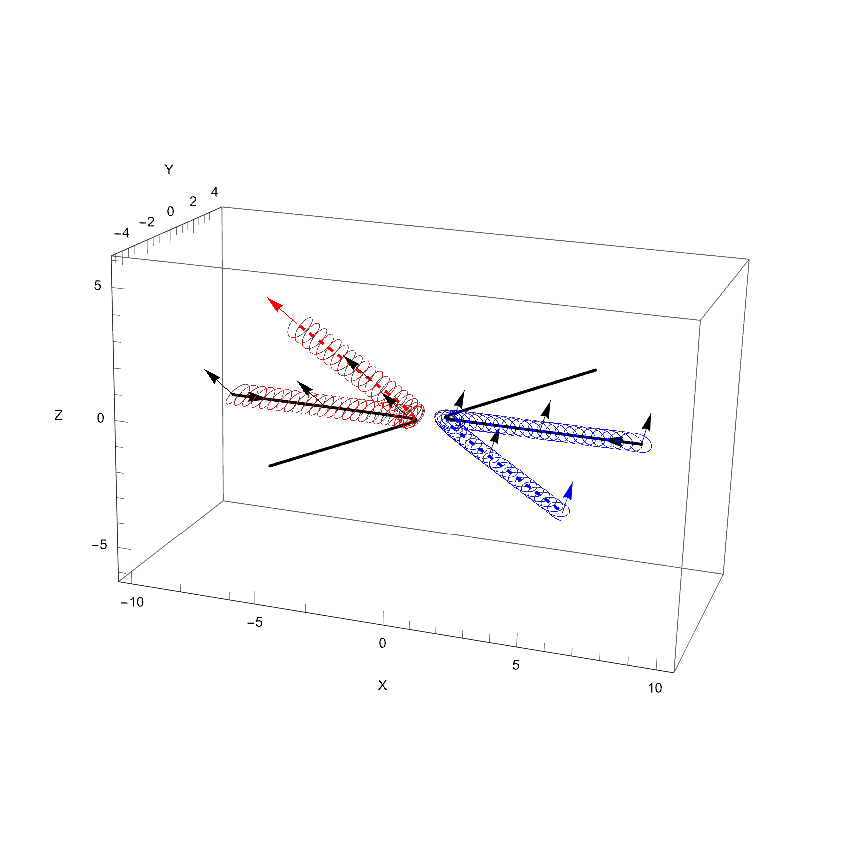}
\includegraphics[scale=0.35]{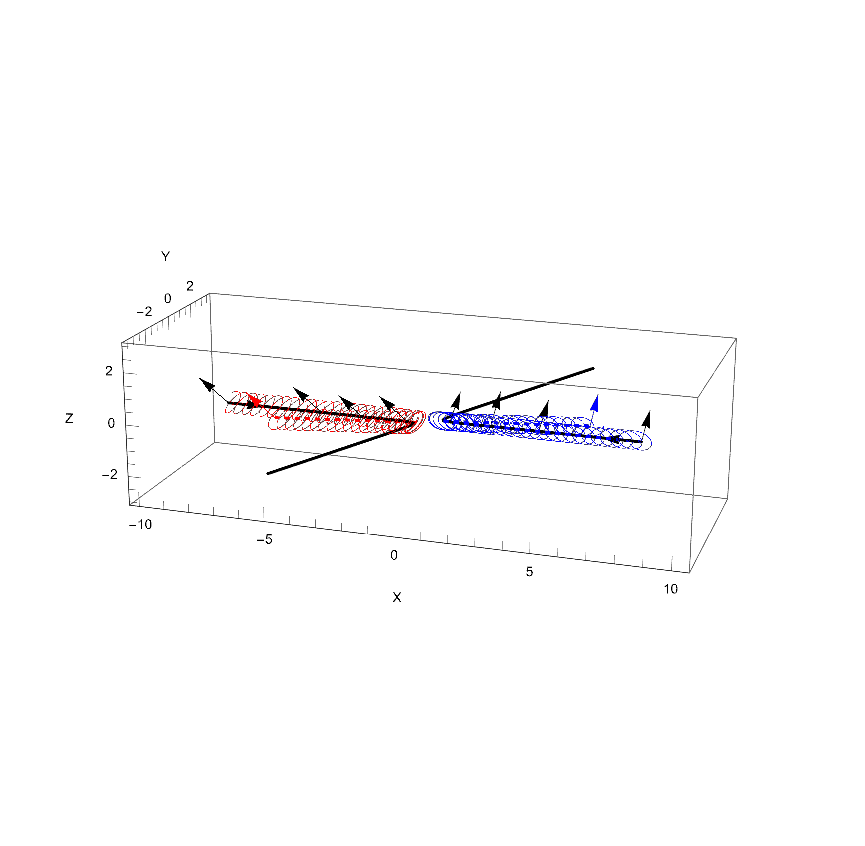}
\includegraphics[scale=0.3]{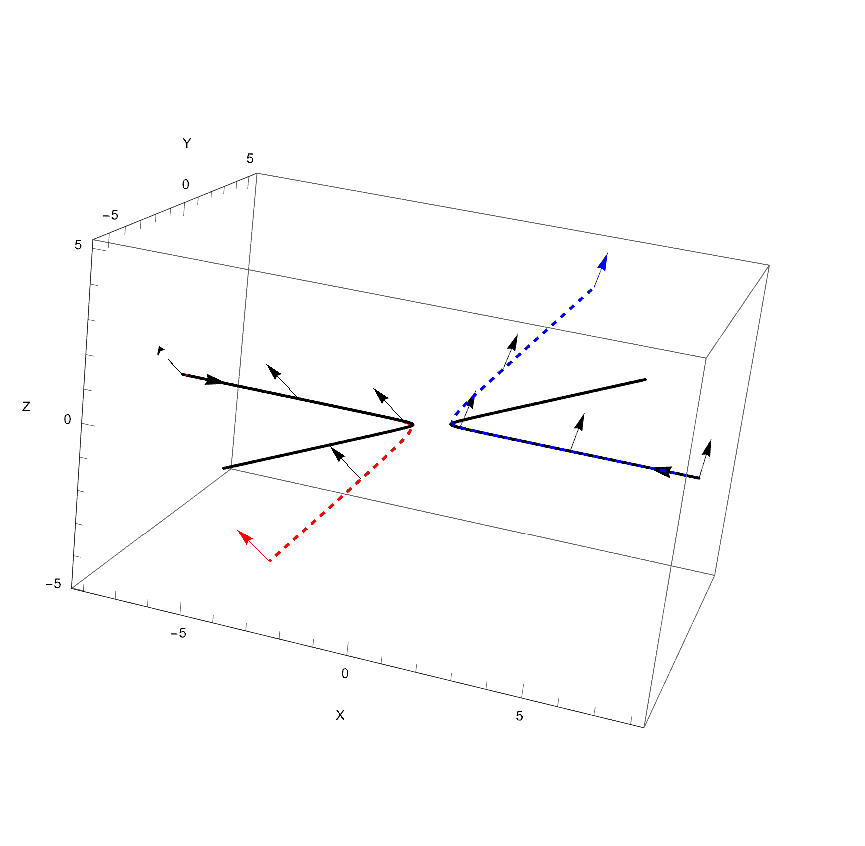}
\includegraphics[scale=0.35]{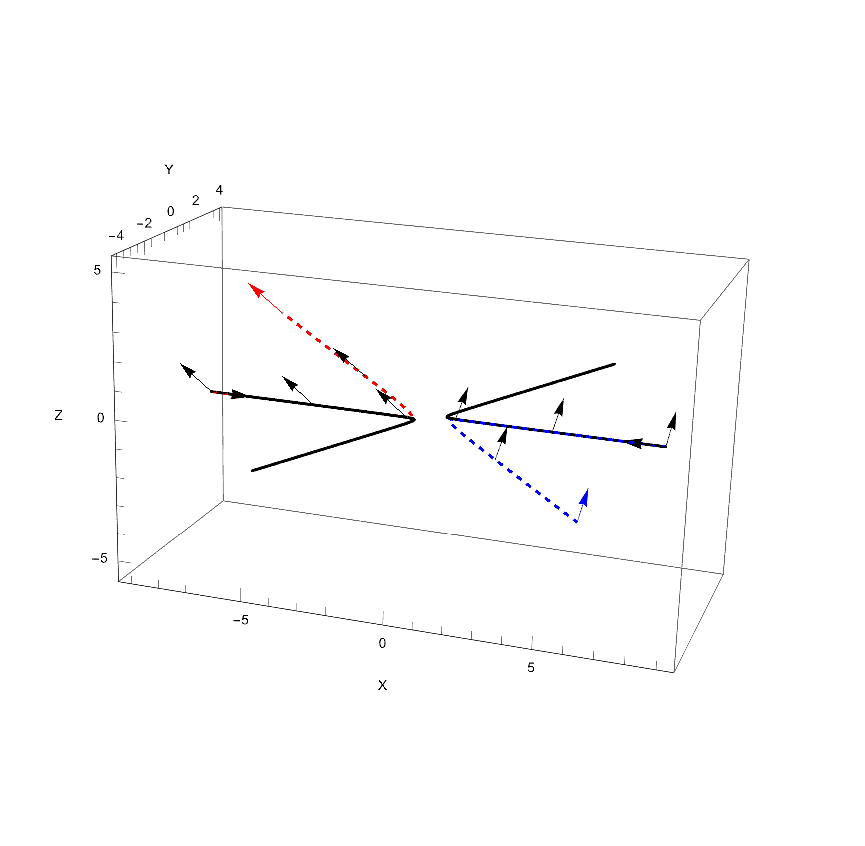}
\includegraphics[scale=0.35]{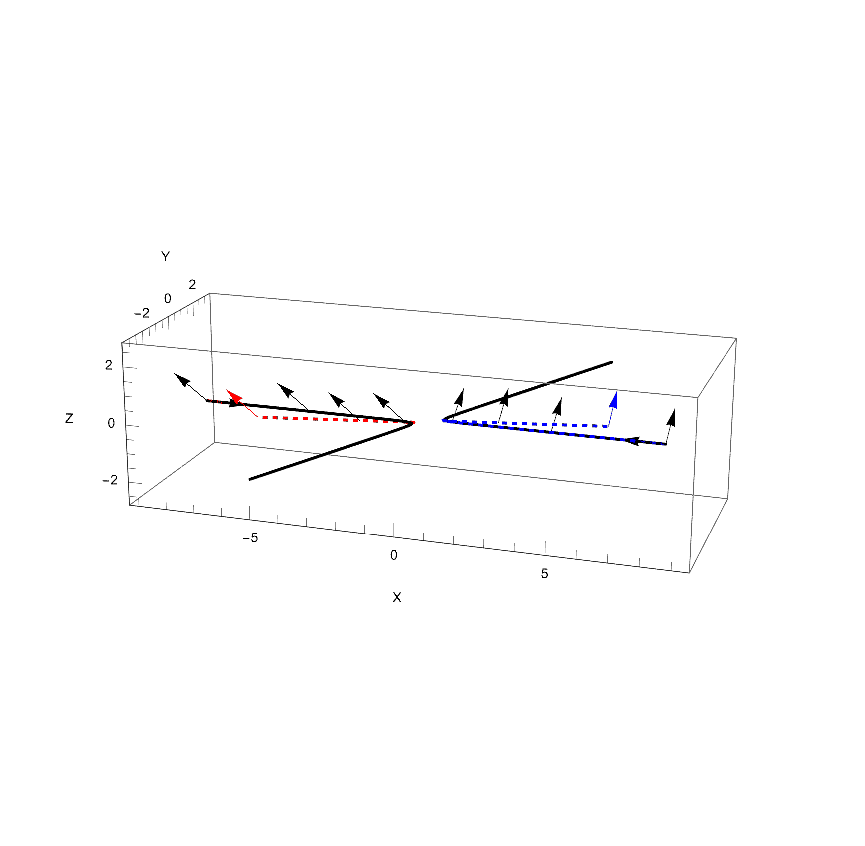}
\includegraphics[scale=0.35]{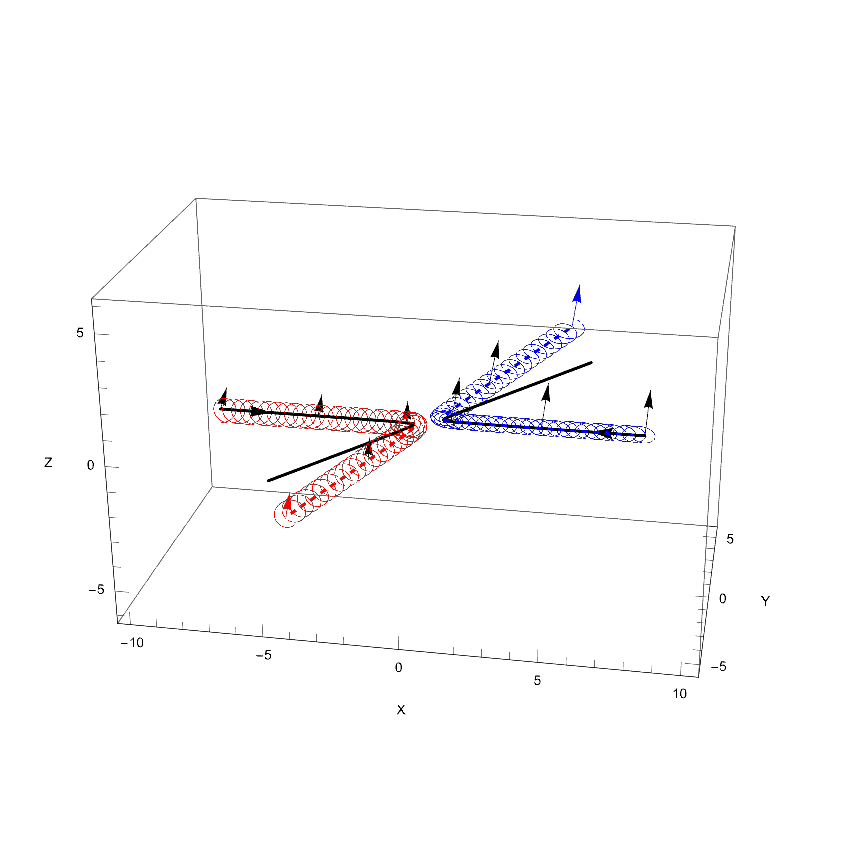}
\includegraphics[scale=0.35]{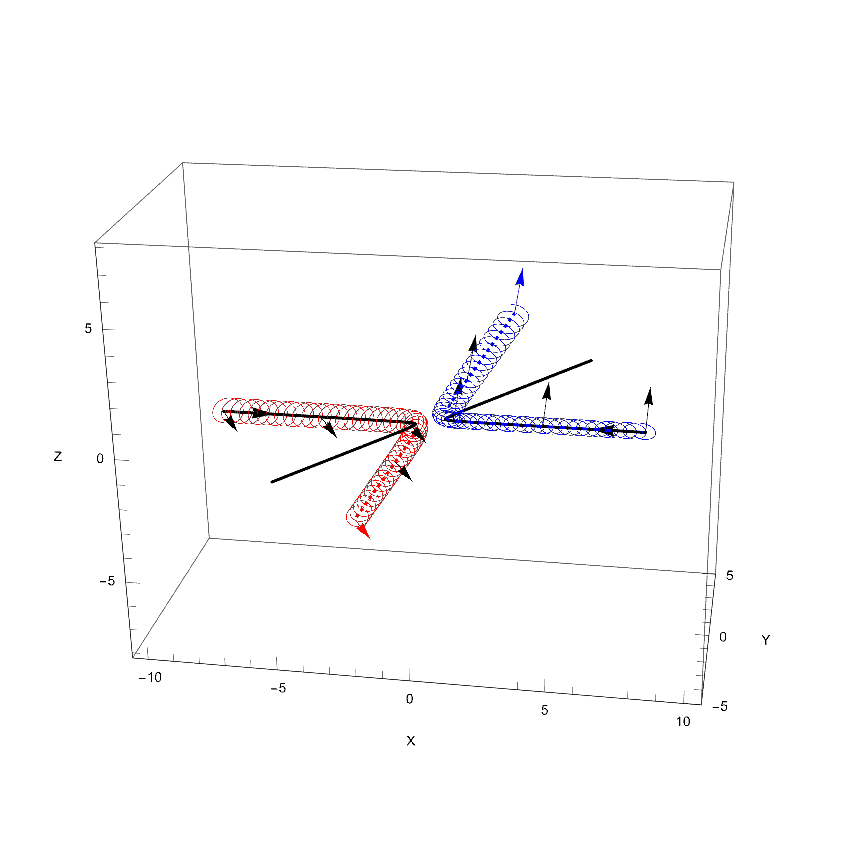}
\includegraphics[scale=0.35]{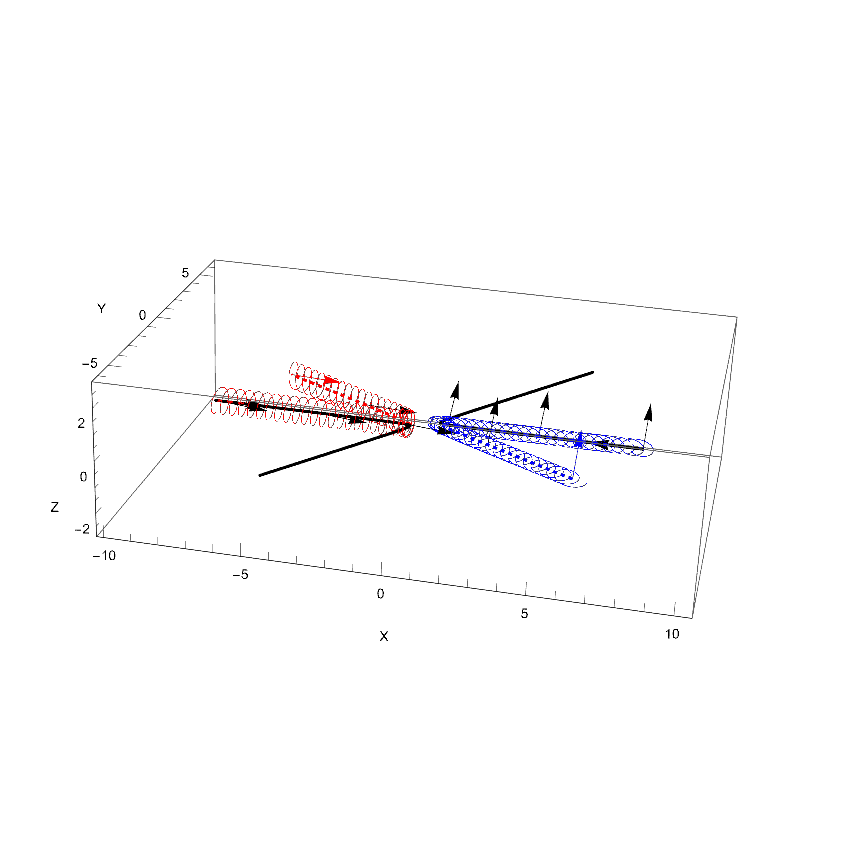}
\caption{Analysis of the Coulomb interaction between two Dirac particles (blue and red), and two point particles (black). A sequence of three integrations with the spins orientations fixed, $\theta_1=60^\circ,\phi_1=90^\circ$ and $\theta_2=45^\circ,\phi_2=220^\circ$, the same CM velocity $v=0.11$ $\beta=90^\circ,\lambda=180^\circ$ and the same initial position of the CM's given by $dx=8,dy=0.01,dz=0.1$. The three figures of the first row correspond to the integration with the phase $\psi_1=60^\circ$ fixed and consecutive values of the phase $\psi_2$ of $0^\circ, 90^\circ$, and $230^\circ$. The figures of the second row are the same integration but without depicting the CC's trajectories.
The third row corresponds to the particle 1 (blue) with the same boundary conditions as in the previous integrations while for particle 2 the spin orientation has been changed to $\theta_2=90^\circ$ and $\phi_2$ takes the values $100^\circ$,  $300^\circ$, and $350^\circ$.
In all cases the Coulomb interaction between the point particles are the black trajectories which are exactly the same in all pictures. We see how a simple modification of the phase which determines the initial position of the CC of Dirac particle 2, $\psi_2$,  or the spin orientation $\phi_2$, we obtain with the same Coulomb force, completely different scattering processes than in the case of the point particles interaction. The relative position of the CC's in the interaction area is crucial for the outcome of the scattering, which is controlled by the spin orientation and the initial phases of the two particles.}
\label{variosfijos}
\end{figure}

In figure {\bf\ref{variosfijos}} we have analyzed the Coulomb interaction between two Dirac particles and also two spinless point particles in which the only boundary variable that is modified is in each integration is the phase $\psi_2$ of one of the Dirac particles. In all pictures the trajectories of the point particles are depicted in black and are the same in all cases, because the position and velocities of the CM's of the particles are the same, and the spin orientations are also the same. We have also fixed the phase $\psi_1=60^\circ$ of the CC of particle 1 and each integral is obtained by changing the CC phase $\psi_2$, which takes the values $0^\circ, 90^\circ$, and $230^\circ$, respectively. In the three integrations the trajectories of the point particles are exactly the same.

In the case of the Coulomb interaction of the point particles, mechanical linear momentum conservation implies that the trajectories of the CM's of both point particles are contained in a plane. The same thing happens for the CM's trajectories of the Dirac particles, but this plane is different than the point particles plane.

Very high energy interaction has not been considered in this preliminary presentation of this subject. The time of one turn of the CC at the center of mass frame is $\pi$ in natural units, and it is $\gamma(v)\pi$ if the CM moves at the speed $v$. An electron of energy 1.14 GeV moves at a velocity of its CM of value $v=0.9999999$, and during a turn of the CC the CM runs a distance $d=\gamma(v)\pi v=14049.6$ in natural units. In a picture at this scale the transversal motion of the CC, of size 1, is almost unobservable and the CC trajectory looks like a straight line, almost parallel to the CM trajectory. To make a drawing we would have to rescale the transversal variables. But we can depict the spin dynamics.

The formalism is relativistic and therefore there is no restriction on the value of the maximum energy allowed for the analysis.

\section{Analysis of the pairing}
\label{Pairing}

In the Coulomb and Poincar\'e case, when the spins are parallel, the two particles are very close to each other with a separation of order of Compton's wavelength or shorter, and the motions of the CC's take place in the same plane we obtain a stable bound motion of the two particles which 
remains stable under external electric fields but which under a transversal magnetic field with respect to the spin direction, is destroyed and the two Dirac particles separate.

\begin{figure}[!hbtp]\centering%
\includegraphics[scale=0.6]{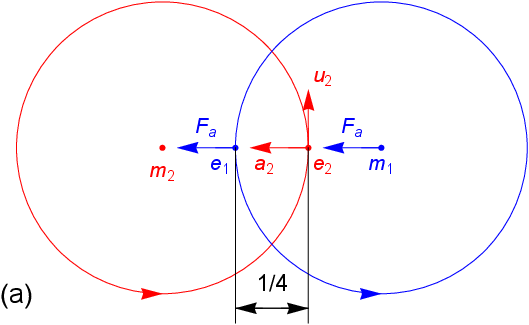}\hspace{1cm}\includegraphics[scale=0.67]{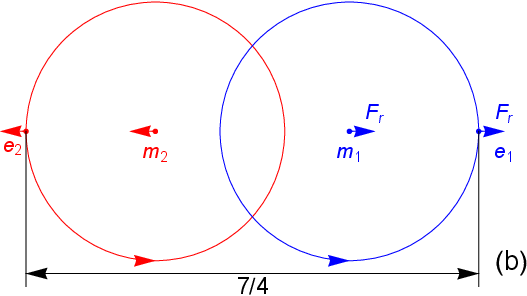}
\caption{Motion of the CC's of the two Dirac particles with parallel spins, a phase shift of the CC's, $\psi_1-\psi_2=180^\circ$, and a separation of their CM's of $3/4$, in natural units. In part (a) the CC's are closer, the repulsive Coulomb force on particle 1 becomes an atractive force $F_a$ between the CM's. After half a turn of the charges, in part (b) the repulsive force between charges is also a repulsive force between their CM's $F_r$ but the intensity of this force is smaller $F_r\ll F_a$. In the Coulomb case  $F_a=49 F_r$, while in the Poincar\'e interaction is greater of value $F_a=343 F_r$, in this example. The two Dirac particles form a stable bound state of spin 1.}
\label{pairingC}
\end{figure}

The justification of this bound motion is because a repulsive force between the CC's can become an atractive force between the CM's as is shown in the figure {\bf\ref{pairingC}-(a)}. Let us analyze the example depicted in the figure where the phases of the two particles are opposite $\psi_1-\psi_2=180^\circ$, and the separation between the CM's is $3/4$ in natural units. Remember that Compton's wavelength is $2R_0$, which is 1 in natural units. In part (a) of the figure the CC's are separated $1/4$ while in part (b), after a half turn of the charges, they are at a distance $7/4$.

We analyze first the Coulomb interaction between the charges and compute the force acting on particle 1. The force on particle 2 is opposite and is not depicted in the figure. In part (a) the repulsive Coulomb force (\ref{eq:CoulombForce}) between the charges is an atractive force $F^C_a$ between the CM's of magnitude in natural units
\begin{equation}
F^C_a=\frac{\alpha}{(1/4)^2}=16\alpha.
\label{coulatrac}
\end{equation}

In part (b) the repulsive Coulomb force between the charges is also a repulsive force $F^C_r$ between the CM's of magnitude in natural units
\[
F^C_r=\frac{\alpha}{(7/4)^2}=\frac{16\alpha}{49}.
\]
But the ratio between the atractive force and the repulsive force in these two points is
\begin{equation}
\frac{F^C_a}{F^C_r}=49.
\label{CoulombRatio}
\end{equation}
The atractive force is 49 times stronger than the repulsive force when the two CC's are far away. In the intermediate situations we have to consider also the direction of the force but when the charges are closer there is always a stronger atractive force than when they are far away. The global result is that the two negative charges form a stable bound state of spin 1.

In the case of the Poincar\'e invariant interaction we obtain the same result but now this ratio is bigger. The Poincar\'e force (\ref{fuerzaPoincare1}) has 4 terms. We analyze this force in the same situation of figure {\bf\ref{pairingC}}. This atractive force in figure {\bf\ref{pairingC}-(a)} between the CM's has a first contribution along ${\bi r}_1-{\bi r}_2$, of intensity
\[
F^P_a(1)=\frac{\alpha\sqrt{2}}{(1/4)^2}=16\alpha\sqrt{2},
\]
because the term in the expression of the force $\sqrt{1-{\bi u}_1\cdot{\bi u}_2}=\sqrt{2}$, since the CC's velocities are opposite to each other. The second term of the force (\ref{fuerzaPoincare1}) on particle 1 depends on the acceleration of the particle 2, with a minus sign, $a_2=2$, so that it represents an atractive term between the CC's but a repulsive force between the CM's, of value
\[
F^P_r(2)=\frac{\alpha}{2(1/4)\sqrt{2}}a_2=2\alpha\sqrt{2}.
\]
The other two terms in the direction of the velocity of the other particle $F^P_a(3)=0$ and $F^P_a(4)=0$, vanish in this situation because $({\bi r}_1-{\bi r}_2)\cdot({\bi u}_1-{\bi u}_2)=0$, and these two vectors are orthogonal to each other and in the fourth term ${\bi a}_1\cdot{\bi u}_2=-{\bi a}_1\cdot{\bi u}_1=0$, and the same for ${\bi a}_2\cdot{\bi u}_1=-{\bi a}_2\cdot{\bi u}_2=0$, because the velocity and acceleration of the CC's are always orthogonal.

The net force in situation {\bf\ref{pairingC}-(a)} is an atractive force between the CM's of intensity
\[
F^P_a=F^P_a(1)-F^P_r(2)=14\alpha\sqrt{2}> F^C_a,
\]
greater than in the Coulomb case (\ref{coulatrac}).

In the situation of figure {\bf\ref{pairingC}-(b)} the repulsive force beteen the CM's is
\[
F^P_r(1)=\frac{\alpha\sqrt{2}}{(7/4)^2}=\frac{16\alpha\sqrt{2}}{49}.
\]
The term which depends on the acceleration is atractive between the CM's of value
\[
F^P_a(2)=\frac{\alpha}{2(7/4)\sqrt{2}}a_2=\frac{2\alpha\sqrt{2}}{7}.
\]
The remaining two terms also vanish, so that the total repulsive force is
\[
F^P_r=\frac{16\alpha\sqrt{2}}{49}-\frac{2\alpha\sqrt{2}}{7}=\frac{2\alpha\sqrt{2}}{49}.
\]
The ratio between the atractive force in {\bf\ref{pairingC}-(a)} to the repulsive force in {\bf\ref{pairingC}-(b)} is thus
\begin{equation}
\frac{F^P_a}{F^P_r}=\frac{14\alpha\sqrt{2}}{2\alpha\sqrt{2}/49}=343\gg\frac{F^C_a}{F^C_r},
\label{PoincRatio}
\end{equation}
much stronger than the ratio for the Coulomb interaction (\ref{CoulombRatio}). This justifies that the Poincar\'e formation of bound pairs can be obtained even at greater relative velocities.

\begin{figure}
\includegraphics[scale=0.35]{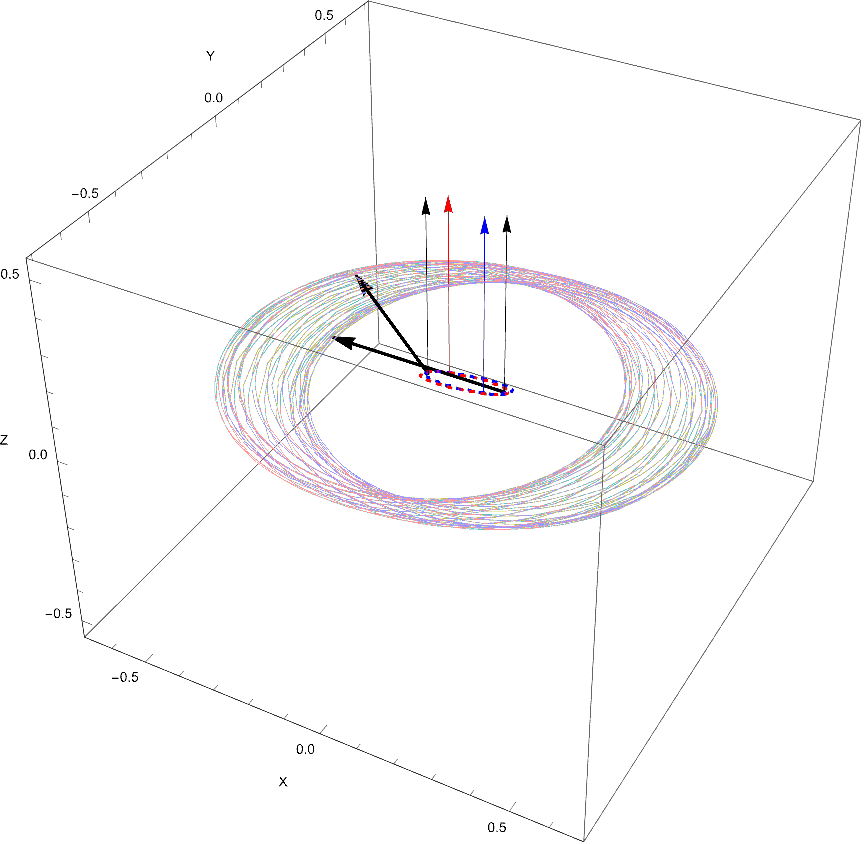}
\includegraphics[scale=0.35]{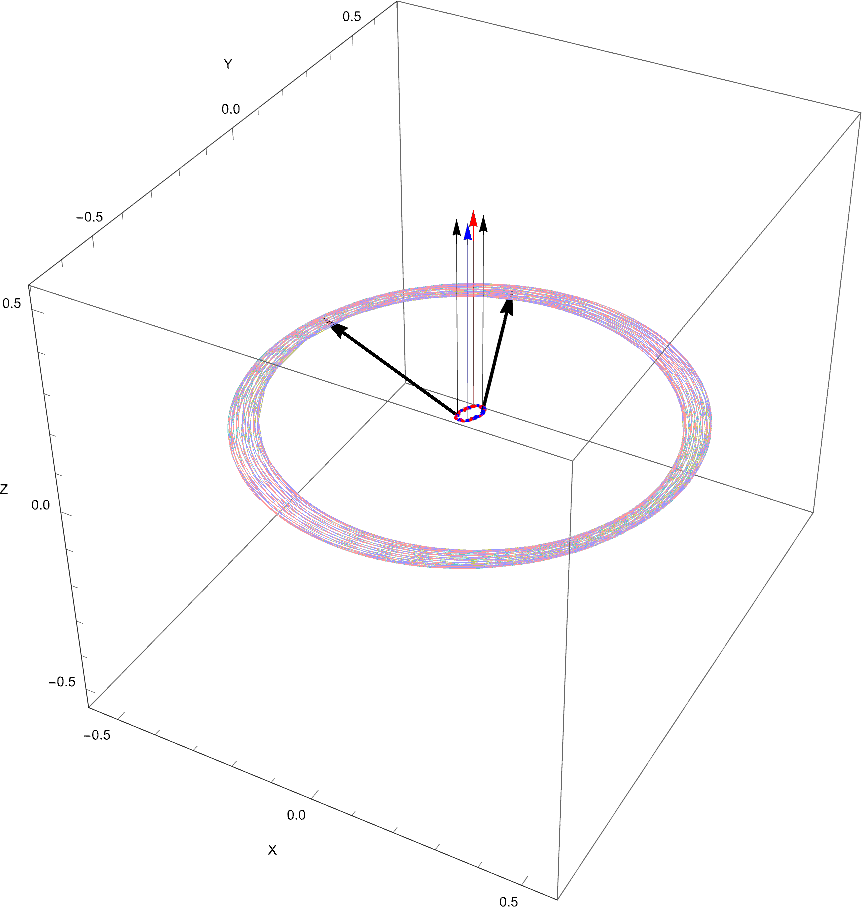}
\includegraphics[scale=0.35]{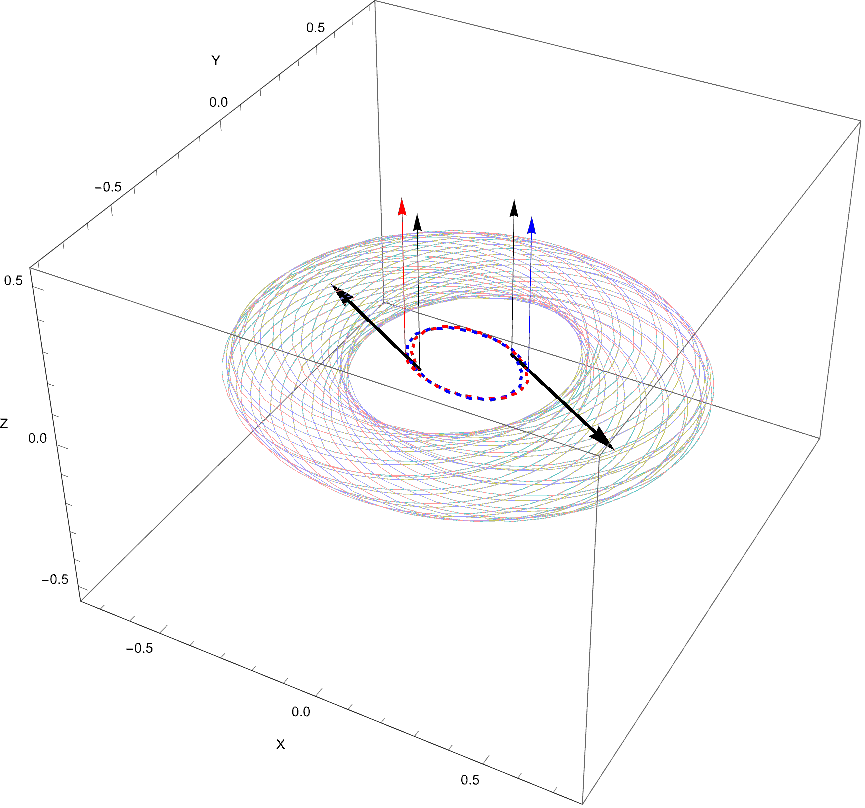}
\caption{Different pairings with $\psi_1-\psi_2\neq180^\circ$, under the Poincar\'e invariant interaction during an integration time of 60 natural units that corresponds to around 20 turns of the CC's. Please remark that the analysis is done in the center of mass frame of the two particles, located at the center of the picture, and the initial velocities of the two particles are not opposite to each other because of the requirement of the total linear momentum conservation. In the case of the Coulomb interaction the initial velocities are opposite to each other because the mechanical momenta are opposite to each other.}
\label{fig:Paired3}
\end{figure}
In the figure {\bf\ref{fig:Paired3}} three different examples of the pairing, during an integration time of 60 in natural units. This corresponds to aroung 20 turns of the CC's and in all of them the phase shift ot the CC's, $\psi_1-\psi_2$, is different than $180^\circ$, where the pairing is more stable. They correspond to $100^\circ$, $100^\circ$ and $140^\circ$, respectively. The initial velocity of the CM's has raised up to $v=0.001$, $0.001$ and $0.002$, and the initial separation of the CM's $dx=0.1, 0.02$, and $0.1$, respectively. Velocities above those values produce the separation of the particles.

This bound state of two electrons is not the classical equivalent of a Cooper pair. Their separation is of the order of the Compton's wavelength, much smaller than the correlation distance of the Cooper pair and it is a spin 1 state. It is thus a bosonic state from the quantum mechanical point of view. Classical physics of spinning particles does not forbid the formation of a stable electron-electron bound state.

The notebook \cite{Slow} describes at a slow motion the pairing of two Dirac particles with parallel spins in the Coulomb interaction.
\section{Conclusions}
\label{Conclusions}
This is not a hidden variables theory. Here all classical variables have a very well defined interpretation like positions, velocities and orientations. The dynamical equations are non-linear and a small modification of the angular variables that define the orientation of the spins or the initial internal phase of the position of the CC's produces a completely different scattering process. Experimental physicists interested in this topic can numerically calculate average values of scattering processes by integration the dynamical equations leaving some of the parameters free and averaging them, for example keeping constant initial spin orientations and averaging over the internal phases $\psi_1$ and $\psi_2$ of the two particles. If the electrons of the beam are unpolarized the average should be done on the $\theta$ and $\phi$ variables of the two particles.


\end{document}